\documentstyle[a4,amstex]{article}

\newcommand{\real}{{\Bbb R}}
\newcommand{\cplx}{{\Bbb C}}
\newcommand{\zint}{{\Bbb Z}}
\newcommand{\cplxn}{\cplx^{n}}
\newcommand{\Ncplxn}{\otimes^N \cplx^{n} }
\newcommand{\cx}{\cplx[x_1 ,x_2 , \dots ,x_N]}
\newcommand{\cz}{\cplx[z_1^{\pm 1},z_2^{\pm 1},\dots,z_N^{\pm 1}]}
\newcommand{\hs}[1]{{\cal H}^{(#1)}}
\newcommand{\h}{{\cal H}}
\newcommand{\ep}{\epsilon}

\newcommand{\gln}{ {{\frak g} {\frak l}}_n}

\newcommand{\Sgroup}[1]{{\frak{S}}_{#1}}

\newcommand{\sprod}[2]{\langle #1, #2 \rangle}
\newcommand{\la}{\lambda}
\renewcommand{\o}{\omega}
\newcommand{\mlett}{m}

\newcommand{\mm}{{\bold m}}
\newcommand{\nn}{{\bold n}}

\newcommand{\MC}{{\cal M}}

\renewcommand{\l}{\lambda}
\newcommand{\s}{\sigma}

\newcommand{\p}{\partial}
\newcommand{\halmos}{\rule{5pt}{5pt}}

\numberwithin{equation}{section}

\newtheorem{prop}{\bf Proposition}
\newtheorem{thm}[prop]{\bf Theorem}

\newenvironment{rmk}{\noindent{\bf Remark}\hskip 5pt}{}

\begin{document}

\begin{titlepage}
\pagestyle{empty}
\begin{center}
\mbox{} \\
\vspace{2.7cm}
\begin{Large}
{\bf The Yangian Symmetry in the Spin Calogero Model and its Applications} 
\end{Large} \\
\vspace{1.5cm}
\large{ Kouichi Takemura }\footnote{e-mail: takemura@@kurims.kyoto-u.ac.jp} 
\\  Research Institute for Mathematical Sciences\\
Kyoto University, Kyoto 606, Japan  \\ 
\vspace{1cm}
February 1997  \\

\vspace{2cm}

\begin{abstract}
By using the non-symmetric Hermite polynomials and a technique based on the Yangian Gelfand-Zetlin bases, we decompose the space of states of the Calogero model with spin into irreducible Yangian modules, construct an orthogonal basis of eigenvectors and derive product-type formulas for norms of these eigenvectors.

\end{abstract}

\end{center}
\end{titlepage}

\section{Introduction.}

 The Calogero model and the Calogero-Sutherland model describe integrable quantum many-body systems with long-ranged interaction \cite{Calogero,Sutherland}.  
 In the Calogero model the particles move on the line $\real ^{1}$,
 and in the Calogero-Sutherland model the particles move on the circle $S^{1}$.
 The spin Calogero-Sutherland model has a non-abelian symmetry identified with the Yangian $Y(\gln )$ \cite{BGHP}.
 In \cite{TU}, the space of states of the spin Calogero-Sutherland model is decomposed into irreducible Yangian submodules and an orthogonal basis of eigenvectors is constructed. Product-type formulas for the norms of these eigenvectors are derived by using the irreducible decomposition of the space of states as the Yangian module and the technique based on the Yangian Gelfand-Zetlin bases \cite{C,nt1,nt2}.

 On the other hand, similarities of the algebraic structures between the spin Calogero-Sutherland model and the spin Calogero model are pointed out in \cite{Kakei,Ujino}. In this paper they are summarized in section \ref{crea}.
Moreover the spin Calogero model has a Yangian symmetry \cite{BHW}.
 In this paper we introduce the Yangian symmetry in the spin Calogero model, in a little different way from \cite{BHW}, decompose the space of states of the spin Calogero model into irreducible Yangian submodules,
 and by using this decomposition construct an orthogonal basis of eigenvectors and derive product-type formulas for the norms of these eigenvectors similar to the spin Calogero-Sutherland model.

 The similarity between these two models appears in the expression of the eigenbasis.
The eigenbasis of the Calogero-Sutherland model is written in terms of the non-symmetric Jack polynomials, and the one in the Calogero model is written in terms of the non-symmetric Hermite polynomials.
These two families of polynomials are deeply connected \cite{BF2,BF3}, in particular see the relation (\ref{gh}) in this paper.
This is the key point why we can treat the Calogero model in a way similar to the one applied to the Calogero-Sutherland model in \cite{TU}.

\vspace{.15in}
{\bf Acknowledgment}
The author would like to thank D. Uglov for useful comments and support. Thanks are also due to Professors M. Jimbo, M. Kashiwara and T. Miwa.

\section{Spin Calogero model with harmonic potential.} \label{cal}

Here we will define the Hamiltonian of the Calogero model and the space of states.

Introduce the Calogero Hamiltonian with spin as follows \cite{Calogero}.
\begin{equation}
H_{CH}:= - h^{2} H_{C} + \omega ^{2} \sum_{j=1}^{N} x_j ^2 , \; \; \; \; 
H_{C}:= \sum_{j=1}^{N} \frac{\partial ^{2}}{\partial x_{j}^{2}}
- \sum _{1 \leq j \neq k \leq N}
\frac{\l ^{2} - \kappa \l P_{j,k}}{(x_{j}-x_{k})^{2}},
\end{equation}

The Hamiltonian $H_{CH}$ acts on wave functions of the form 
$ \phi (x_{1}, \dots , x_{N} \; | \; s_{1}, \dots , s_{N}) $, 
 where $s_{i} \; (1 \leq i \leq N)$ takes values in 
$ \{ 1, \dots , n \}$ and is called the spin variable.\\
 The operator $P_{i,j} $ acts on the wave functions and changes the $i$-th spin
and the $j$-th spin:
\begin{equation}
 (P_{i,j}  \phi) ( \dots | \dots , s_{i} , \dots , s_{j} , \dots )
:=  \phi ( \dots | \dots , s_{j} , \dots , s_{i} , \dots ) .
\end{equation} 
The constant $\kappa $ is $\pm 1$ and the system is called bosonic (resp. fermionic) if $\kappa =1$ (resp. $=-1$).
The wavefunction of the bosonic (resp. fermionic) Calogero model is symmetric (resp. antisymmetric) with respect to the exchangement of the coordinate and the spin at the same time.

\begin{rmk}
For the spinless bosonic Calogero model,
 the wave function of the ground state is as follows:
\begin{equation}
\tilde{\phi }_0 = e^{- \frac{\omega }{2h} \sum_{j=1}^N x_j ^2 } 
\prod_{j<k} |x_j -x_k | ^{\l } .
\end{equation}
\end{rmk}

 We introduce the following gauge transformation and a new Hamiltonian,
\begin{equation}
 \phi _0 := \prod_{j<k} |x_j -x_k | ^{\l }, 
\end{equation}
\vspace{-.2in}
\begin{equation}
\tilde{H}_{C}:= \phi _0^{-1} H_{C} \phi _0= 
 \sum_{j=1}^{N} \frac{\partial ^{2}}{\partial x_{j}^{2}} 
+ \sum _{1 \leq j \neq k \leq N} \left\{
 \frac{2\l }{x_j -x_k} \frac{\partial }{\partial x_{j}} 
+ \frac{\l (\kappa P_{i,j} -1)}{(x_j -x_k)^{2}} \right\}, 
\end{equation}
\vspace{-.2in}
\begin{equation}
\tilde{H}_{CH} := \phi _0^{-1} H_{CH} \phi _0 =- h^2 \tilde{H}_{C} +
  \omega ^{2} \sum_{j=1}^{N} x_j ^2.
\label{CH}
\end{equation}
The spaces of states of the gauge transformed bosonic $(\kappa = 1)$ (resp. fermionic $(\kappa =-1)$) Calogero model are 
\begin{equation}
\hs{\kappa}:= \bigcap_{i=1}^{N-1} Ker(K_{i,i+1}P_{i,i+1} - \kappa 1) \subset  
{\cal H} ( := e^{- \frac{\omega }{2h} \sum_{j=1}^N x_j ^2 } \cdot
\cx \otimes(\Ncplxn))
\end{equation}
Here $K_{i,j}$ exchanges $x_{i}$ and $x_{j}$,
and $P_{i,j}$ exchanges the $i$--th component and $j$--th component of 
$ \Ncplxn$.

We will define the scalar product on the space $\cal{H}^{(\kappa )}$.
For this purpose we first define  the scalar product on $\Ncplxn$ and 
$e^{- \frac{\omega }{2h} \sum_{j=1}^N x_j ^2 } \cdot \cx $.

We fix the base $\{v_{\ep}\}_{\ep=1,\dots,n}$ in $\cplxn$ and define in $\Ncplxn$ the hermitian (sesquilinear) scalar product ${\sprod{\;\cdot\;}{\;\cdot\;}}_s$ by requiring pure tensors to be orthonormal: 
\begin{equation}
{\sprod{v_{\ep_1}\otimes v_{\ep_2}\otimes \cdots \otimes v_{\ep_N}}{v_{\tau_1}\otimes v_{\tau_2}\otimes \cdots \otimes v_{\tau_N}}}_s:= \prod_{i=1}^N \delta_{\ep_i,\tau_i}\qquad (\ep_i,\tau_i = 1,2,\dots,n). 
\label{sps}
\end{equation}
In $e^{- \frac{\omega }{2h} \sum_{j=1}^N x_j ^2 } \cdot \cx (\ni f,g)$ define 
the scalar product ${\sprod{\;\cdot\;}{\;\cdot\;}}_{c}$ as follows.
\begin{equation}
{\sprod{f}{g}}_c := \left( \prod _{i=1}^{N} \int_{-\infty}^{\infty} dx_{i}
\right) \prod_{j<k} | x_{j}-x_{k} |^{2\l }
\overline{f(x_1,x_2,\dots,x_N)} g(x_1,x_2,\dots,x_N).
\label{sph}
\end{equation}
The hermitian scalar product ${\sprod{\;\cdot\;}{\;\cdot\;}}$ in the space $\h$ is defined as the composition of the scalar products (\ref{sps}) and (\ref{sph}). For $f, g$ $\in$ $e^{- \frac{\omega }{2h} \sum_{j=1}^N x_j ^2 } \cdot \cx$
 and $u, v$ $\in$ $\Ncplxn$ put   
\begin{equation}
\sprod{f\otimes u}{g\otimes v} := {\sprod{f}{g}}_c{\sprod{u}{v}}_s 
\end{equation}
and extend the ${\sprod{\;\cdot\;}{\;\cdot\;}}$ on the entire space $\h$ by requiring it to be sesquilinear.
We define the scalar product 
${\sprod{\;\cdot\;}{\;\cdot\;}}_{(\kappa )}$
 on the space $\cal{H}^{(\kappa )}$ by
restricting the scalar product ${\sprod{\;\cdot\;}{\;\cdot\;}}$ to the subspace $\cal{H}^{(\kappa )}$.

\section{Degenerate affine Hecke algebra and the non-symmetric Jack polynomial.}

In this section we recall the definition of the non-symmetric Jack polynomial.

We define the operator $d_{i} \; (1 \leq i \leq N)$ which act on 
$\cplx [ x_{1}, x_{2}, \dots x_{N} ] $ as follows.
\begin{equation}
d_{i}:= \frac{\p }{\p x_{i}} + \l \sum_{j \neq i}
\frac{1-K_{i,j}}{x_i - x_j }. 
\end{equation}
Then we can check following relations.
\begin{equation}
[ d_i , d_j ] =0, \; \; \; K_{i,j}d_i = d_j K_{i,j}
\end{equation}
\begin{rmk}
If $v \in {\cal H}^{(\kappa )}$ then we have
\begin{equation}
\left(
\sum_{i=1}^{N} d_{i}^2 \right) v = \tilde{H}_{C} \: v. \; \; \; \; 
\end{equation}
\end{rmk}
We put $\alpha = 1/\l $ and set
\begin{equation}
\tilde{d}_{i} := \alpha x_i d_i + \sum_{j>i} K_{i,j}, \; \; \; 
\hat{d}_i :=\tilde{d}_{i} - N.
\end{equation}
Then $\hat{d}_i$ are the Dunkl operators which appear in the Calogero-Sutherland model \cite{Dunkl,BGHP}. (See \cite{} (2.9))
\begin{equation}
\hat{d}_i = \alpha \frac{\p }{\p x_{i}} -i +
\sum_{j >i} \frac{x_{j}}{x_j -x_i} (K_{i,j}- 1) -
\sum_{j <i} \frac{x_{i}}{x_i -x_j} (K_{i,j}- 1) .
\end{equation}
Remark that the operators $\tilde{d}_{i}$ and $K_{i,i+1}$ satisfy
 the relations of the degenerate affine Hecke algebra.
\begin{eqnarray}
& [ \tilde{d}_{i} ,\tilde{d}_{j}] =0 ,& \\
& K_{i,i+1} \tilde{d}_{i} -\tilde{d}_{i+1} K_{i,i+1} = 1 . & \\
& [\tilde{d}_{i} ,K_{j,j+1}] =0 ,& |i-j| >1.
\end{eqnarray}
The actions of $\tilde{d}_{i}$ on monomials are as follows.
\begin{equation}
\tilde{d}_{i} \cdot x_1 ^{n_{1}} x_2 ^{n_{2}} \dots x_N ^{n_{N}} =
(\alpha n_i -i+N) x_1 ^{n_{1}} x_2 ^{n_{2}} \dots x_N ^{n_{N}} 
\label{daction}
\end{equation}
\vspace{-.2in}
\begin{equation}
+\sum_{j<i} \left\{
\begin{array}{cc}
\sum_{l=n_j +1}^{n_i } (x_{i}^{l} x_j ^{n_i + n_j -l} 
\prod _{m \neq i,j} x_{m}^{n_m}) &
(n_i > n_j) \\
0 &  (n_i = n_j) \\
\sum_{l=n_i +1}^{n_j } - (x_{i}^{l} x_j ^{n_i + n_j -l} 
\prod _{m \neq i,j} x_{m}^{n_m}) &
(n_i < n_j)  
\end{array}
\right.
\nonumber
\end{equation}
\vspace{-.2in}
\begin{equation}
- \sum_{j>i} \left\{
\begin{array}{cc}
\sum_{l=n_j }^{n_i -1} (x_{i}^{l} x_j ^{n_i + n_j -l} 
\prod _{m \neq i,j} x_{m}^{n_m}) &
(n_i > n_j) \\
0 &  (n_i = n_j) \\
\sum_{l=n_i }^{n_j -1} - (x_{i}^{l} x_j ^{n_i + n_j -l} 
\prod _{m \neq i,j} x_{m}^{n_m}) &
(n_i < n_j)  
\end{array}
\right.
.
\nonumber
\end{equation}
Let $\MC _N := $ $\{ (m_1,m_2,\dots,m_N) \in \zint^N \; | \; m_1 \geq 
m_2 \geq \dots \geq m_N \geq 0 \}$ be the set of partitions.
For $\mm \in \MC _N$ we set
\begin{equation}
S^{\mm } := \{ \sigma \in {\frak{S}}_{N} | i<j \mbox{ and }m_{\s i}=m_{\s j} \Rightarrow \s (i) < \s (j) \}.
\label{sm}
\end{equation}
We can identify the pair $(\mm , \sigma ) \; \; (\mm : \mbox{ partition } \; ,
\s \in S^{\mm })$ and $\nn ( \in \zint_{\geq 0}^N)$ as follows.
\begin{equation}
(\mm , \sigma ) \Leftrightarrow 
\nn = (m_{\s (1)} ,m_{\s (2)} ,\dots , m_{\s (N)} ).
\end{equation} 
Then $\s (\in S^{\mm })$ satisfy the following property:
\begin{equation}
\text{ for all $1\leq i \leq N $} \quad \s(i) = \#\{\;j\leq i \;| \; m_{\s(j)} \geq m_{\s(i)}\} 
+ \#\{\;j>i \;| \; m_{\s(j)} >  m_{\s(i)}\}.  
\end{equation}

In the set $S^{\mm}$  introduce the total ordering by setting
\begin{gather}
\s \succ \s' , \quad \quad \text{iff } \quad \text{the last non-zero element} \label{eq:Sord} \\ \text{ in $(\mlett_{\s(1)} - \mlett_{\s'(1)},\mlett_{\s(2)}-\mlett_{\s'(2)},\dots,\mlett_{\s(N)}-\mlett_{\s'(N)})$ is $< 0$.} \nonumber
\end{gather}
Notice that the identity in $\Sgroup{N}$ is the maximal element in $S^{\mm}$ in this ordering. 
Then in the set of pairs $(\mm,\s)$ $(\mm \in \MC_N, \; \s \in S^{\mm})$ the partial ordering is defined by 
\begin{equation}
(\mm,\s) > (\tilde{\mm},\tilde{\s}) \quad \text{iff} \quad \begin{cases} \mm > \tilde{\mm} & \text{or} \\
\mm = \tilde{\mm}, &  \s \succ \tilde{\s} \end{cases}
\end{equation}
where $\mm > \tilde{\mm}$ means that $\mm$ is greater than $\tilde{\mm}$ in the dominance (natural) ordering in $\MC_N$,
\begin{equation}
\mm > \tilde{\mm} \; \; \; \; \; \Leftrightarrow 
\end{equation}
\vspace{-.15in}
\begin{equation}
\mm \neq \tilde{\mm } , \; \; \; |\mm |:= \sum_{i=1}^{N} m_{i} =|\tilde{\mm }| ,\; \; \mbox{ and }
\sum_{i=1}^{j} m_i \geq \sum_{i=1}^{j} \tilde{m}_i \; \;
 (1 \leq \forall j \leq N-1).
\nonumber
\end{equation}

The eigenvectors $\Phi_{\s}^{\mm}(x) \in \cx$ of the Dunkl operators are labeled by the pairs $(\mm,\s)$ $(\mm \in \MC_N, \; \s \in S^{\mm})$ and satisfy the following properties: 
\begin{align}
& \Phi_{\s}^{\mm}(z) = x_1^{\mlett_{\s(1)}}x_2^{\mlett_{\s(2)}}\cdots x_N^{\mlett_{\s(N)}} + \sum_{(\tilde{\mm},\tilde{\s}) < ({\mm},{\s})} c_{({\mm},{\s});(\tilde{\mm},\tilde{\s})} x_1^{\tilde{\mlett}_{\tilde{\s}(1)}}x_2^{\tilde{\mlett}_{\tilde{\s}(2)}}\cdots x_N^{\tilde{\mlett}_{\tilde{\s}(N)}}; \\  
& \tilde{d}_i \Phi_{\s}^{\mm}(x) = \xi_i^{\mm}(\s) \Phi_{\s}^{\mm}(x), \quad \text{where}\quad \xi_i^{\mm}(\s):= \alpha \mlett_{\s(i)} +N - \s(i) \qquad (i=1,2,\dots,N); \label{xi} \\
& K_{i,i+1}\Phi_{\s}^{\mm}(x) = {\cal{A}}_i^{\mm}(\s)\Phi_{\s}^{\mm}(x) + {\cal{B}}_i^{\mm}(\s)\Phi_{\s(i,i+1)}^{\mm}(x),  \\ 
\intertext{where}
& {\cal{A}}_i^{\mm}(\s) := \frac{1}{ \xi_i^{\mm}(\s) -  \xi_{i+1}^{\mm}(\s)},  \label{AB} \\ &  {\cal{B}}_i^{\mm}(\s) := \begin{cases} \frac{\left(\xi_i^{\mm}(\s) -  \xi_{i+1}^{\mm}(\s)\right)^2 - 1}{\left(\xi_i^{\mm}(\s) -  \xi_{i+1}^{\mm}(\s)\right)^2}  &
(\mlett_{\s(i)} >  \mlett_{\s(i+1)}) , \\ \qquad  0  &  (\mlett_{\s(i)} =  \mlett_{\s(i+1)}), \\
\qquad 1 & (\mlett_{\s(i)} <  \mlett_{\s(i+1)}). \end{cases} 
\nonumber
\end{align}
Notice that for $\s \in S^{\mm}$ we have $\s(i+1) = \s(i)+1$ whenever $\mlett_{\s(i)} =  \mlett_{\s(i+1)}$, and hence in this case, 
\begin{equation}
K_{i,i+1}\Phi_{\s}^{\mm}(x) = \Phi_{\s}^{\mm}(x) \qquad (\mlett_{\s(i)} =  \mlett_{\s(i+1)}). 
\end{equation}
We remark that if $\alpha >0$ then the jointeigenvalues are all distinct and $\Phi_{\s}^{\mm}(x)$ are well-defined.
We call these polynomials $\Phi_{\s}^{\mm}(x)$ the non-symmetric Jack polynomials \cite{Cherednik2,Macdonald1}.
If we set $\psi := x_N K_{N-1,N} K_{N-2,N-1} \dots K_{1,2} $, we get (c.f. \cite{Knop})
\begin{equation}
\psi \cdot \Phi^{\nn }(x) = \Phi^{\nn '}(x). 
\end{equation}
Here, $ \Phi^{(m_{\s (1)}, \dots ,m_{\s (N)})} := \Phi_{\s}^{\mm}(x) $,
$\nn \; (\in \zint ^{N}_{\geq 0}) := (n_1, n_2, \dots n_N )$, and
  $\nn ' :=  (n_2, n_3, \dots n_N , n_1 +1)$.

For $f(x_1,x_2,\dots,x_N)$, $g(x_1,x_2,\dots,x_N)$ $\in$ $\cx$ set 
\begin{equation}  
{\sprod{f}{g}}_J := \frac{1}{N!}\left( \prod_{i=1}^N \oint_{|w_i|=1} \frac{dw_i}{2\pi \sqrt{-1}w_i}  \right) \left( \prod_{ i\neq j} 1 - \frac{w_i}{w_j} \right)^{\frac{1}{\alpha}} \overline{f(w_1,w_2,\dots,w_N)} g(w_1,w_2,\dots,w_N)  
\end{equation}
where the integration  over each of the complex variables $w_i$ is taken along the unit circle in the complex plane. 

If we denote by $A^{\dagger }$ the adjoint of the operator $A$ with respect to the scalar product  ${\sprod{\;\cdot\;}{\;\cdot\;}}_{J}$, we get
\begin{equation}
K_{i,i+1} ^{\dagger }= K_{i,i+1}, \; \; \; x_{i} ^{\dagger }= x_{i}^{-1},
\end{equation}
\vspace{-.2in}
\begin{equation}
\tilde{d}_{i} ^{\dagger }=  \tilde{d}_{i}, \; \; \; 
\psi ^{\dagger }= \psi ^{-1}.
\end{equation}
The non-symmetric Jack polynomials are orthogonal with respect to
 ${\sprod{\;\cdot\;}{\;\cdot\;}}_{J}$ and the recursive relations for the norms are
\begin{equation}
(1-{\cal{A}}_i^{\mm}(\s)^{2}) 
{\sprod{\Phi_{\s}^{\mm}(x) }{\Phi_{\s}^{\mm}(x) }}_{J}
={\cal{B}}_i^{\mm}(\s)^{2}
{\sprod{\Phi_{\s (i,i+1)}^{\mm}(x) }{\Phi_{\s (i,i+1)}^{\mm}(x) }}_{J},
\label{recjack1}
\end{equation}
\vspace{-.2in}
\begin{equation}
{\sprod{\Phi^{\nn}(x) }{\Phi^{\nn}(x)}}_{J}=
{\sprod{\Phi^{\nn '}(x) }{\Phi^{\nn '}(x)}}_{J}.
\label{recjack2}
\end{equation}
where $\nn \; (\in \zint ^{N}_{\geq 0}) := (n_1, n_2, \dots n_N )$, and
  $\nn ' :=  (n_2, n_3, \dots n_N , n_1 +1)$.

\section{Creation operators and the non-symmetric generalized Hermite polynomials.} \label{crea}

We will introduce the creation (annihilation) operators, the Dunkl operators for the Calogero model, and the non-symmetric generalized Hermite polynomials. These polynomials are also introduced in \cite{BF2,BF3} in a little different way.

In this section, we will deal with the following scalar product (\ref{sph}):
\begin{equation}
{\sprod{f}{g}}_c := \left( \prod _{i=1}^{N} \int_{-\infty}^{\infty} dx_{i}
\right) \prod_{j<k} | x_{j}-x_{k} |^{2\l }
\overline{f(x_1,x_2,\dots,x_N)} g(x_1,x_2,\dots,x_N).
\nonumber
\end{equation}

First we can check that
\begin{equation}
d_{i}^{\dagger }= -d_{i}, \; \; \; \; x_{i}^{\dagger }= x_{i}.
\end{equation}
We define the annihilation operators as follows,
\begin{equation}
A_{i} := hd_i + \omega x_{i} , \; \; \; \; 
\bar{A}_i := \frac{1}{2h\omega }A_{i} . 
\end{equation}
Then the adjoints of the operators $A_{i}$ are
\begin{equation}
A_{i}^{\dagger} := - hd_i + \omega x_{i}.
\end{equation}
We call $A_{i}^{\dagger}$ the creation operators. 
The commutation relations are
\begin{eqnarray}
& [A_{i}^{\dagger}, A_{j}^{\dagger}]= [ \bar{A}_i, \bar{A}_j ]=0 , & \label{ca1} \\
& [ A_{i}^{\dagger}, K_{j,k}] = [ \bar{A}_{i}, K_{j,k}]=0 ,&   (i\neq j,k),\label{ca2} \\ 
& A_{i}^{\dagger} K_{i,j} = K_{i,j} A_{j}^{\dagger} , \; \; \; \;
\bar{A}_{i} K_{i,j} = K_{i,j} \bar{A}_{j}, & \label{ca3} \\
& [ \bar{A}_{i}, A_{j}^{\dagger} ] = \delta _{i,j} \left( 
1 + \l \sum_{k \neq i} K_{i,k} \right) + (1-\delta_{i,j} )\l K_{i,j} .&
\label{ca4}
\end{eqnarray}
\begin{rmk}
If we replace $A_{i}^{\dagger} \rightarrow  x_{i}, \; \; \; 
\bar{A}_{i} \rightarrow d_i $, the same relations as (\ref{ca1} -- \ref{ca4})
are satisfied \cite{Kakei,Ujino}.
\end{rmk}
\vspace{.15in}

We introduce the Dunkl operators 
\begin{equation}
\tilde{\Delta }_{i} := \alpha A_{i}^{\dagger} \bar{A}_{i} 
+ \sum_{j>i} K_{i,j}, \; \; \; \; (\alpha = \frac{1}{\l }).
\end{equation}
Then the operators $\tilde{\Delta }_{i}$ and $K_{i,i+1}$ satisfy
 the relations of the degenerate affine Hecke algebra.
\begin{eqnarray}
& [ \tilde{\Delta }_{i} ,\tilde{\Delta }_{j}] =0 ,& \label{CdH1} \\
& K_{i,i+1} \tilde{\Delta }_{i} -\tilde{\Delta }_{i+1} K_{i,i+1} = 1 , & \label{CdH2} \\
& [\tilde{\Delta }_{i} ,K_{j,j+1}] =0 ,& |i-j| >1.\label{CdH3} 
\end{eqnarray}
The sum of the Dunkl operators is essentially the same as
 the Hamiltonian $\tilde{H}_{CH}$.
\begin{equation}
\left( \sum_{i=1}^{N} \tilde{\Delta }_{i} \right) v = 
\left( \frac{-1}{2h \omega \l } \tilde{H}_{CH} + \frac{N}{2 \l } 
\right) v , \; \; \; \; \; \mbox{ if } v \in {\cal H}^{(\kappa )}.
\end{equation}

Next, we will calculate the joint eigenfunctions of the operators
 $\tilde{\Delta }_{i}$. we will use the following relations later.
\begin{equation}
\bar{A}_{i} e^{ -\frac{\omega }{2h} \sum_{j=1}^N x_j ^2 } =0, \; \; \; \;
K_{i,k} e^{ -\frac{\omega }{2h} \sum_{j=1}^N x_j ^2 } =
 e^{ -\frac{\omega }{2h} \sum_{j=1}^N x_j ^2 }.
\label{ani}
\end{equation}
If we move $\bar{A}_{i}$ and $K_{i,j}$ to the right by the relations (\ref{ca2},\ref{ca3},\ref{ca4}) and use the relations (\ref{ani}), we get
\begin{equation}
\tilde{\Delta }_{i} (A_{1}^{\dagger}) ^{n_{1}} (A_{2}^{\dagger} )^{n_{2}}
 \dots (A_{N}^{\dagger}) ^{n_{N}}e^{ -\frac{\omega }{2h} 
\sum_{j=1}^N x_j ^2 }
\end{equation}
\vspace{-.2in}
\begin{equation} 
=(\alpha n_i -i+N) (A_{1}^{\dagger}) ^{n_{1}} (A_{2}^{\dagger}) ^{n_{2}} \dots 
(A_{N}^{\dagger}) ^{n_{N}} e^{ -\frac{\omega }{2h} \sum_{j=1}^N x_j ^2 } 
\nonumber
\end{equation}
\vspace{-.2in}
\begin{equation}
+\sum_{j<i} \left\{
\begin{array}{cc}
\sum_{l=n_j +1}^{n_i } (A_{i}^{\dagger})^{l} (A_{j}^{\dagger}) ^{n_i + n_j -l} 
\prod _{m \neq i,j} (A_{m}^{\dagger})_{m}^{n_m} &
(n_i > n_j) \\
0 &  (n_i = n_j) \\
\sum_{l=n_i +1}^{n_j } - (A_{i}^{\dagger})^{l} (A_{j}^{\dagger}) ^{n_i + n_j -l} 
\prod _{m \neq i,j} (A_{m}^{\dagger})^{n_m} &
(n_i < n_j)  
\end{array}
\right\}
e^{ -\frac{\omega }{2h} \sum_{j=1}^N x_j ^2 }
\nonumber
\end{equation}
\vspace{-.2in}
\begin{equation}
- \sum_{j>i} \left\{
\begin{array}{cc}
\sum_{l=n_j }^{n_i -1} (A_{i}^{\dagger})^{l} (A_{j}^{\dagger}) ^{n_i + n_j -l} 
\prod _{m \neq i,j} (A_{m}^{\dagger})^{n_m} &
(n_i > n_j) \\
0 &  (n_i = n_j) \\
\sum_{l=n_i }^{n_j -1} -(A_{i}^{\dagger})^{l} (A_{j}^{\dagger})^{n_i + n_j -l} 
\prod _{m \neq i,j} (A_{m}^{\dagger})^{n_m} &
(n_i < n_j)  
\end{array}
\right\}
e^{ -\frac{\omega }{2h} \sum_{j=1}^N x_j ^2 }
.
\nonumber
\end{equation}
Comparing to the relation (\ref{daction}), the coefficients of 
$x_1 ^{k_1} \dots x_N ^{k_N}$ and $(A_{1}^{\dagger})^{k_1}\dots (A_{N}^{\dagger})^{k_N} e^{ -\frac{\omega }{2h} \sum_{j=1}^N x_j ^2 }$
are the same for all $(k_1 , \dots ,k_N)$. So the joint eigenfunction
of the operators $\tilde{\Delta }_{i}$ are written by using the non-symmetric
 Jack polynomials:
\begin{equation}
\tilde{\Delta }_{i} \Phi_{\s}^{\mm}(A_{1}^{\dagger}, \dots ,A_{N}^{\dagger})
e^{ -\frac{\omega }{2h} \sum_{j=1}^N x_j ^2 }
= \xi_i^{\mm}(\s) \Phi_{\s}^{\mm}(A_{1}^{\dagger}, \dots ,A_{N}^{\dagger})
e^{ -\frac{\omega }{2h} \sum_{j=1}^N x_j ^2 },
\end{equation}
\vspace{-.2in}
\begin{equation}
 \text{where}\quad \xi_i^{\mm}(\s):= \alpha \mlett_{\s(i)} +N - \s(i) \qquad (i=1,2,\dots,N).
\nonumber
\end{equation}
Because of the relations 
\begin{eqnarray}
& [ A_{i}^{\dagger} ,A_{j}^{\dagger} ] = [ x_i , x_j ] =0, & \\
& [ A_{i}^{\dagger} ,K_{j,k} ] = [ x_i ,K_{j,k} ]=0 , & (i\neq j,k) \\
& A_{i}^{\dagger} K_{i,j} = K_{i,j} A_{j}^{\dagger}, \; \; \; \; 
x_i K_{i,j} = K_{i,j} x_j, & 
\end{eqnarray}
we get
\begin{equation}
K_{i,i+1}\Phi_{\s}^{\mm}(A_{1}^{\dagger}, \dots ,A_{N}^{\dagger})
e^{ -\frac{\omega }{2h} \sum_{j=1}^N x_j ^2 } 
\label{Kact}
\end{equation}
\vspace{-.2in}
\begin{equation}
= {\cal{A}}_i^{\mm}(\s)\Phi_{\s}^{\mm}(A_{1}^{\dagger}, \dots ,A_{N}^{\dagger})
e^{ -\frac{\omega }{2h} \sum_{j=1}^N x_j ^2 }
 + {\cal{B}}_i^{\mm}(\s)\Phi_{\s(i,i+1)}^{\mm}(A_{1}^{\dagger}, \dots ,A_{N}^{\dagger})e^{ -\frac{\omega }{2h} \sum_{j=1}^N x_j ^2 },
\nonumber              
\end{equation}
\vspace{-.2in}
\begin{equation}
\tilde{\psi } \cdot \Phi^{\nn }(A_{1}^{\dagger}, \dots ,A_{N}^{\dagger}) e^{ -\frac{\omega }{2h} \sum_{j=1}^N x_j ^2 }
\label{psiact}
\end{equation}
\vspace{-.2in}
\begin{equation}
:= A_{N}^{\dagger} K_{N-1,N} K_{N-2,N-1} \dots K_{1,2} \Phi^{\nn }(A_{1}^{\dagger}, \dots ,A_{N}^{\dagger})e^{ -\frac{\omega }{2h} \sum_{j=1}^N x_j ^2 }  
\nonumber
\end{equation}
\vspace{-.2in}
\begin{equation}
= \Phi^{\nn '}(A_{1}^{\dagger}, \dots ,A_{N}^{\dagger})
e^{ -\frac{\omega }{2h} \sum_{j=1}^N x_j ^2 }.
\nonumber
\end{equation}
Here  $(\mm \in \MC_N, \; \s \in S^{\mm})$ , $\nn \; (\in \zint ^{N}_{\geq 0}) = (n_1, n_2, \dots n_N )$, and
  $\nn ' =  (n_2, n_3, \dots n_N , n_1 +1)$.
 The coefficients $\cal{A}_i^{\mm}(\s), \; \; \cal{B}_i^{\mm}(\s)$
 are the same as those in the Jack case (\ref{AB}).

We define the non-symmetric generalized Hermite polynomials as follows.
\begin{equation}
\Phi_{\s}^{\mm (H)}(x_{1} , \dots ,x_{N})
:= e^{ \frac{\omega }{2h} \sum_{j=1}^N x_j ^2 } 
\left( \frac{1}{2\omega } \right) ^{| \mm |} 
\Phi_{\s}^{\mm}(A_{1}^{\dagger}, \dots ,A_{N}^{\dagger})
e^{ -\frac{\omega }{2h} \sum_{j=1}^N x_j ^2 }
\label{gh}
\end{equation}
If we set the degrees as follows,
\begin{equation}
deg(\omega ):=-2, \; \; \; \; deg(h):=0 , \; \; \; \; deg(x_{i}):=1,
\end{equation}
the polynomial $\Phi_{\s}^{\mm (H)}(x_{1} , \dots ,x_{N})$ is homogeneous
and $deg (\Phi_{\s}^{\mm (H)}(x_{1} , \dots ,x_{N}))= |\mm |$.
The non-symmetric generalized Hermite polynomials have the following expansion:
\begin{equation}
 \Phi_{\s}^{\mm (H)}(x) =  \Phi_{\s}^{\mm }(x)
+ \sum_{|\nn | < |\mm |} c ^{\mm , \: \s}_{\nn }
x_{1}^{n_1} x_{2}^{n_2} \dots x_{N}^{n_N}.
\end{equation}
We note the relationship between the symmetric Jack (resp. Hermite) polynomials
and the non-symmetric Jack (resp. Hermite) polynomials.
We define the symmetric Jack (resp. Hermite) polynomials as the 
joint eigenfunctions for the operators 
$\sum_{i=1}^{N}\tilde{d}_{k}^{i} $ 
(resp. $e^{ \frac{\omega }{2h} \sum_{j=1}^N x_j ^2 } \cdot \sum_{i=1}^{N}
( \hat{\Delta }_{k}^{i}) \cdot e^{ -\frac{\omega }{2h} \sum_{j=1}^N x_j ^2 } $) $(k=1 , \dots , N)$.
They are labeled by the partitions and normalized so that the coefficient of
the highest term (for the dominance ordering of the partitions) is $1$
\cite{Stanley,BF1}.
The symmetric Jack (resp. Hermite) polynomials are written by the sum of 
the non-symmetric Jack (resp. Hermite) polynomials as follows.
\begin{equation}
J^{\mm }(x) = \sum_{\s \in S^{\mm}} c_{\s } \Phi_{\s}^{\mm }(x), \; \; \; \; \; (\mbox{resp. }H^{\mm }(x) = \sum_{\s \in S^{\mm}} c_{\s } \Phi_{\s}^{\mm (H)}(x)).
\label{symnons}
\end{equation}
Remark that the coefficients $c_{\s }$ for the Jack case and for the Hermite
 case are the same.

We calculate the recursion relations for the norms of 
$\Phi_{\s}^{\mm}(A_{1}^{\dagger}, \dots ,A_{N}^{\dagger})
e^{ -\frac{\omega }{2h} \sum_{j=1}^N x_j ^2 }$. \\
From the relation (\ref{Kact}) we get
\begin{equation}
(1-{\cal{A}}_i^{\mm}(\s)^{2}) 
{\sprod{\Phi_{\s}^{\mm}(A_{1}^{\dagger}, \dots ,A_{N}^{\dagger}) 
e^{ -\frac{\omega }{2h} \sum_{j=1}^N x_j ^2 } 
}{\Phi_{\s}^{\mm}(A_{1}^{\dagger}, \dots ,A_{N}^{\dagger})
e^{ -\frac{\omega }{2h} \sum_{j=1}^N x_j ^2 }  }}_{c}
\label{hAB}
\end{equation}
\vspace{-.2in}
\begin{equation}
={\cal{B}}_i^{\mm}(\s)^{2}
{\sprod{\Phi_{\s (i,i+1)}^{\mm}(A_{1}^{\dagger}, \dots ,A_{N}^{\dagger})
e^{ -\frac{\omega }{2h} \sum_{j=1}^N x_j ^2 }  }{\Phi_{\s (i,i+1)}^{\mm}(A_{1}^{\dagger}, \dots ,A_{N}^{\dagger})
e^{ -\frac{\omega }{2h} \sum_{j=1}^N x_j ^2 }  }}_{c}.
\nonumber
\end{equation}
If we put $\mm : \mbox{ partition } \; ,\s \in S^{\mm }), $
 $\; \nn =(m_{\s (1)} ,\dots , m_{\s (N)})$, and
 $\nn ' =  (n_2, n_3, \dots n_N , n_1 +1)$,
 by using the relation (\ref{psiact}) we get 
\begin{equation}
{\sprod{\Phi^{\nn '}(A_{1}^{\dagger}, \dots ,A_{N}^{\dagger})
e^{ -\frac{\omega }{2h} \sum_{j=1}^N x_j ^2 }
  }{\Phi^{\nn '}(A_{1}^{\dagger}, \dots ,A_{N}^{\dagger})
e^{ -\frac{\omega }{2h} \sum_{j=1}^N x_j ^2 } }}_{c}= \dots 
\label{hnn}
\end{equation}
\vspace{-.2in}
\begin{equation}
= \frac{2h\omega}{\alpha } 
{\sprod{\Phi^{\nn }(A_{1}^{\dagger}, \dots ,A_{N}^{\dagger})
e^{ -\frac{\omega }{2h} \sum_{j=1}^N x_j ^2 }
  }{(\tilde{\Delta }_{1} + \alpha )
\Phi^{\nn }(A_{1}^{\dagger}, \dots ,A_{N}^{\dagger})
e^{ -\frac{\omega }{2h} \sum_{j=1}^N x_j ^2 } }}_{c}
\nonumber
\end{equation}
\vspace{-.2in}
\begin{equation}
= 
2h\omega 
\left( m_{\s (1)}+1 +\frac{N- \s (1)}{\alpha } \right)
{\sprod{\Phi^{\nn }(A_{1}^{\dagger}, \dots ,A_{N}^{\dagger})
e^{ -\frac{\omega }{2h} \sum_{j=1}^N x_j ^2 }
  }{\Phi^{\nn }(A_{1}^{\dagger}, \dots ,A_{N}^{\dagger})
e^{ -\frac{\omega }{2h} \sum_{j=1}^N x_j ^2 } }}_{c}.
\nonumber
\end{equation}
By comparing to the Jack cases (\ref{recjack1},\ref{recjack2}), we have 
\begin{equation}
\frac{{\sprod{\Phi_{\s}^{\mm}(A_{1}^{\dagger}, \dots ,A_{N}^{\dagger})
e^{ -\frac{\omega }{2h} \sum_{j=1}^N x_j ^2 }  
}{\Phi_{\s}^{\mm}(A_{1}^{\dagger}, \dots ,A_{N}^{\dagger})
e^{ -\frac{\omega }{2h} \sum_{j=1}^N x_j ^2 }  }}_{c}}{
{\sprod{\Phi_{\s (i,i+1)}^{\mm}(A_{1}^{\dagger}, \dots ,A_{N}^{\dagger})
e^{ -\frac{\omega }{2h} \sum_{j=1}^N x_j ^2 }  
}{\Phi_{\s (i,i+1)}^{\mm}(A_{1}^{\dagger}, \dots ,A_{N}^{\dagger}) 
e^{ -\frac{\omega }{2h} \sum_{j=1}^N x_j ^2 } }}_{c}} 
\label{ratio1}
\end{equation}
\vspace{-.2in}
\begin{equation}
 =
\frac{{\sprod{\Phi_{\s}^{\mm}(x) }{\Phi_{\s}^{\mm}(x) }}_{J}
}{{\sprod{\Phi_{\s (i,i+1)}^{\mm}(x) }{\Phi_{\s (i,i+1)}^{\mm}(x) }}_{J}},
\nonumber
\end{equation}
\vspace{-.2in}
\begin{equation}
\frac{{\sprod{\Phi^{\nn '}(A_{1}^{\dagger}, \dots ,A_{N}^{\dagger})
e^{ -\frac{\omega }{2h} \sum_{j=1}^N x_j ^2 }  
}{\Phi^{\nn '}(A_{1}^{\dagger}, \dots ,A_{N}^{\dagger})
e^{ -\frac{\omega }{2h} \sum_{j=1}^N x_j ^2 }  }}_{c}}{
{\sprod{\Phi^{\nn }(A_{1}^{\dagger}, \dots ,A_{N}^{\dagger})
e^{ -\frac{\omega }{2h} \sum_{j=1}^N x_j ^2 }  
}{\Phi^{\nn }(A_{1}^{\dagger}, \dots ,A_{N}^{\dagger})
e^{ -\frac{\omega }{2h} \sum_{j=1}^N x_j ^2 }  }}_{c}} 
\label{ratio2}
\end{equation}
\vspace{-.2in}
\begin{equation}
= 2h\omega
\left( m_{\s (1)} +1+\frac{N -\s (1)}{\alpha } \right)
\frac{{\sprod{\Phi^{\nn '}(x) }{\Phi^{\nn '}(x)}}_{J}}{
{\sprod{\Phi^{\nn }(x) }{\Phi^{\nn }(x)}}_{J}}.
\nonumber
\end{equation}
Because of the relations (\ref{symnons},\ref{ratio1},\ref{ratio2}), we can calculate the norms of the symmetric Hermite polynomials.
\begin{equation}
{\sprod{H^{\mm }(x) e^{ -\frac{\omega }{2h} \sum_{j=1}^N x_j ^2 }}{
H^{\mm }(x)e^{ -\frac{\omega }{2h} \sum_{j=1}^N x_j ^2 }}}_{c}
\label{normsh}
\end{equation}
\vspace{-.2in}
\begin{equation}
=
\left( \frac{\omega }{2h} \right) ^{|\mm |}
\prod_{(i,j) \in \mm } \left( j+\frac{N-i}{\alpha } \right)
{\sprod{J^{\mm }(x) }{J^{\mm }(x)}}_{J}
\frac{ {\sprod{ e^{ -\frac{\omega }{2h} \sum_{j=1}^N x_j ^2 }}{
e^{ -\frac{\omega }{2h} \sum_{j=1}^N x_j ^2 }}}_{c} }{
{\sprod{1 }{1}}_{J} }
\nonumber
\end{equation}
\vspace{-.2in}
\begin{equation}
= \left( \frac{\omega }{2h} \right) ^{|\mm |+ \frac{N(N-1)}{2\alpha }}
 \left( \frac{\omega \pi}{h} \right) ^{\frac{N}{2}} 
N ! 
\prod _{i=1}^{N} \Gamma (m_i +1 +\frac{N-i}{\alpha } )
\nonumber
\end{equation}
\vspace{-.2in}
\begin{equation} 
\cdot \prod _{1 \leq i<j \leq N}
\frac{\Gamma \left( m_{i} - m_{j} + \frac{-i+j+1}{\alpha } \right)
\Gamma \left( m_{i} - m_{j} +1+ \frac{-i+j-1}{\alpha } \right)}{
\Gamma \left( m_{i} - m_{j} +1+ \frac{-i+j}{\alpha } \right)
\Gamma \left( m_{i} - m_{j} + \frac{-i+j}{\alpha } \right)}.
\nonumber
\end{equation}
$\prod _{(i,j) \in \mm }$ means the product over the $(i,j)$--boxes contained in the Young diagram $\mm $. To get the formula (\ref{normsh}), we used the following formulas \cite{Macdonaldbook} chap.VI 10.38, \cite{BF1} prop 3.7.
\begin{equation}
{\sprod{J^{\mm }(x) }{J^{\mm }(x)}}_{J}=
\prod _{1 \leq i<j \leq N }
\frac{\Gamma \left( m_{i} - m_{j} + \frac{-i+j+1}{\alpha } \right)
\Gamma \left( m_{i} - m_{j} +1+ \frac{-i+j-1}{\alpha } \right) }{
\Gamma \left( m_{i} - m_{j} +1+ \frac{-i+j}{\alpha } \right)
\Gamma \left( m_{i} - m_{j} + \frac{-i+j}{\alpha } \right) },
\end{equation}
\vspace{-.2in}
\begin{equation}
\left( \prod _{i=1}^{N} \int_{-\infty}^{\infty} 
e^{-x_{i}^{2}} dx_{i}
\right) \prod_{j<k} | x_{j}-x_{k} |^{\frac{2}{\alpha } }
= 2^{- \frac{N(N-1)}{2\alpha }} \pi ^{\frac{N}{2}} 
\prod _{j=1}^{N-1} \frac{\Gamma (1+ \frac{j+1}{\alpha })}{\Gamma (1+\frac{1}{\alpha })},
\end{equation}
\vspace{-.2in}
\begin{equation}
\left( \prod_{i=1}^N \oint_{|w_i|=1} \frac{dw_i}
{2\pi \sqrt{-1}w_i}  \right) \left( \prod_{ i\neq j} 1 - \frac{w_i}{w_j} 
\right) ^{\frac{1}{\alpha}}
= \frac{\Gamma (1+ \frac{N}{\alpha })}{\Gamma (1+\frac{1}{\alpha })^N}.
\end{equation}
Remark that if $\frac{\omega }{h} =1$ then the formula (\ref{normsh}) is written in \cite{BF1} prop 3.7.

\section{Yangian $Y(\gln )$ and the Yangian Gelfand-Zetlin bases.} \label{ygh}

In this section we summarize properties of the Yangian $Y(\gln )$ which are used in this paper. The main attention is given to the Gelfand-Zetlin algebra and the canonical Yangian Gelfand-Zetlin bases in certain irreducible Yangian modules.

The Yangian $Y(\gln )$ is a unital associative algebra generated
by the elements $1$ and $T_{a,b}^{(s)}$ where $a,b =1,\cdots,n$ and $s=1,2,\cdots$
that are subject to the following relations: 
\begin{equation}
[T_{a,b}^{(r)},T_{c,d}^{(s+1)}] - [T_{a,b}^{(r+1)},T_{c,d}^{(s)}] = T_{c,b}^{(r)}T_{a,d}^{(s)} - T_{c,b}^{(s)}T_{a,d}^{(r)} \qquad ( r,s = 0,1,2,\dots \;) 
\label{Ydr}
\end{equation}
where  $ T_{a,b}^{(0)} := \delta_{a,b}1$. 

Introducing the formal Taylor series in
$u^{-1}$
\begin{equation}
T_{a,b}(u)=\delta _{a,b} +T_{a,b}^{(1)}u^{-1}
+T_{a,b}^{(2)} u^{-2}+\ldots .
\end{equation}
Define $\stackrel{k}{T}(u) \: (k=1,2)$ as follows.
\begin{equation}
\stackrel{k}{T}(u)=\sum_{a,b=1}^{n} E_{a,b}^{(k)} \otimes T_{a,b}(u)
 \in End(\cplxn ) \otimes End(\cplxn ) \otimes Y(\gln )[[u^{-1}]].
\end{equation}
Here $E_{a,b}^{(k)}$ are the standard matrix units that are acting in the $k$-th tensor factor $\cplxn$.
If we put 
\begin{equation}
R(u,v)= 1 + \frac{1}{u-v} \sum_{a,b=1}^{n}  E_{a,b}^{(1)} 
\otimes E_{b,a}^{(2)}
\end{equation}
then the defining relations of $Y(\gln )$ are
\begin{equation}
R(u,v) \stackrel{1}{T} (u) \stackrel{2}{T}(v)
 = \stackrel{2}{T}(v) \stackrel{1}{T}(u) R(u,v).
\label{RTT}
\end{equation}

 Let
${\bold i}=(i_1,\dots,i_m)$ and ${\bold j}=(j_1,\dots,j_m)$ be two
sequences of indices such that
\begin{equation}
1\leq i_1<\ldots<i_m\leq n
\; \mbox{ and } \;
1\leq j_1<\ldots<j_m\leq n.
\end{equation}
Let ${{\frak S} _{m}}$ be the symmetric group of degree $m$. Define  
\begin{equation}
Q_{{\bold i}{\bold j}}(u)=
\sum_{\sigma \in {{\frak S} _{m}}}
sgn(\sigma ) \cdot T_{i_1, j_{\sigma (1)}}(u) T_{i_2, j_{\sigma (2)}}(u-1)
\dots T_{i_m, j_{\sigma (m)}}(u-m+1),
\end{equation}
and
\begin{equation}
A_{0}(u)=1, \; \; \; A_{m}(u)= Q_{{\bold i}{\bold i}}(u) , \; \; (m=1, \cdots ,n)
\end{equation}
\begin{equation}
B_{m}(u)= Q_{{\bold i}{\bold j}}(u), \; \; 
C_{m}(u)= Q_{{\bold j}{\bold i}}(u). \; \; 
 ( m=1, \cdots ,n-1)
\end{equation}
 where ${\bold i}=(1,\ldots,m)$ and ${\bold j}=(1,\ldots,m-1,m+1)$.
 The following propositions are can be found in the paper \cite{nt1}.
\begin{prop}  {\em\cite{nt1} } \label{pr1}
a) The coefficients of $A_n(u)$ belong to the center of the algebra
$Y(\gln )$. \\
 b) All the coefficients of $A_1(u),\dots,A_n(u)$ pairwise commute.
\end{prop}

By Proposition \ref{pr1}, the coefficients $ A_m^{(s)} $ of the series
 $A_{1}(u), \dots A_{n}(u)$:
\begin{equation}
A_m(u) = \sum_{s\geq 0} u^{-s} A_m^{(s)} \qquad ( m=1,2,\dots,n)
\end{equation}
 generate the commutative subalgebra in $Y(\gln)$. 
 This algebra is called Gelfand-Zetlin algebra and is denoted by $A(\gln )$.
\vspace{.2in}

Let $V$ be an irreducible finite dimensional $\gln$-module and $E_{a,b}$ be the generators of $\gln$.
 Denote by $v_{\l }$ the  highest weight vector in $V$ :
\begin{equation}
E_{a,a}\cdot v_{\l }=\l_a v_{\l } \qquad E_{a,b}\cdot v_{\l } =0,\quad a<b.
\end{equation}
Then each difference $\l_a-\l_{a+1}$ is a non-negative integer.
 We assume that each $\l_a$ is also an integer.
Denote by ${\cal T}_{\l } $ the set of all arrays $\Lambda$
 with integral entries of the form
\begin{eqnarray}
& \l _{n,1} \; \; \l _{n,2} \; \cdots \cdots \cdots \cdots \cdots \; \; 
 \l _{n,n} &  \\
& \l _{n-1,1} \; \;   \cdots \cdots \; \; \l _{n-1,n-1} &
\nonumber \\
& \ddots \; \; \cdots \cdots \; \; \; \; \; \; \;  & \nonumber \\
& \l _{2,1} \; \; \l _{2,2} & \nonumber \\
& \l _{1,1} & \nonumber
\end{eqnarray}
where $\l _{n,i}=\l _i$ and $\l _i \geq \l _{m,i}$ for all $i$ and $m$.
The array $\Lambda $ is called a  Gelfand-Zetlin scheme if
\begin{equation}
\l_{m,i}\geq\l_{m-1,i}\geq\l_{m,i+1}
\end{equation}
for all possible $m$ and $i$.
 Denote by ${\cal S}_{\l }$ the subset in ${\cal T}_{\l }$
consisting of the Gelfand-Zetlin schemes \cite{gz}.

\vspace{.2in}
Let us recall some facts about  representations of the Yangian $Y(\gln )$.

If we set $u'=u+h, v'=v+h \; \; (h \in \cplx )$, the relations (\ref{RTT})
 are also satisfied for $(u',v')$.
Thus the map
\begin{equation}
T_{a,b}(u) \mapsto T_{a,b}(u+h) 
\end{equation}
defines an automorphism of the algebra $Y(\gln )$.
So if there is a representation $V$ of $Y(\gln )$, we can construct another
representation of $Y(\gln )$ by the pullback through this automorphism.

We can regard the representation of the Lie algebra $\gln $
 as the representation of  $Y(\gln )$. This transpires  due to the existence of the  homomorphism $\pi_n$ from $Y(\gln )$ to $U(\gln )$:
 the universal enveloping algebra of $\gln $:
\begin{equation}
\pi_{n} \; : \; T_{a,b}(u) \mapsto \delta_{a,b} + E_{b,a}u^{-1}.
\label{pin}
\end{equation}
 
 Let $V_{\l }$ be the irreducible $\gln $--module whose highest weight is 
$\l = (\l_{1} , \l_{2} , \dots ,\l_{n})$.
 We denote by $V_{\l}(h)$ the $Y(\gln )$--module obtained from $V_{\l }$
 by the pullback through this homomorphism and the automorphism (\ref{pin}).

\vspace{.2in} 
The Yangian $Y(\gln )$ has the coproduct
 $\Delta :Y(\gln ) \rightarrow Y(\gln ) \otimes Y(\gln )$.
 It is given as follows.
\begin{equation}
\Delta (T_{a,b}(u)) = \sum_{c=1}^{n} T_{a,c}(u) \otimes T_{c,b}(u) .   
\end{equation}
So if there are representations $V_{i} \; (i=1, \dots ,M)$
 of the Yangian $Y(\gln )$,
we can construct the representation $V_{1} \otimes V_{2} \otimes \cdots 
\otimes V_{M}$ of $Y(\gln )$:
\begin{equation}
T_{a,b}(u) \cdot (v_{1} \otimes v_{2} \otimes \cdots \otimes v_{M}) =
\Delta ^{(M)}\circ \cdots \circ \Delta ^{(2)} 
(T_{a,b}(u)) (v_{1} \otimes v_{2} \otimes \dots \otimes v_{M} ) 
\label{copro}
\end{equation}
\vspace{-.15in}
\begin{equation}
= \sum_{k_{1} \dots k_{M-1}} T_{a,k_{1}}(u) v_{1} \otimes 
T_{k_{1},k_{2}}(u) v_{2} \otimes \cdots \otimes T_{k_{M-1},b}(u) v_{M}.
\nonumber
\end{equation}

\mbox{} From now on
 we consider the following representation of the Yangian $Y(\gln )$: 
\begin{equation}
W= V_{\lambda ^{(1)}}(h^{(1)}) \otimes  V_{\lambda ^{(2)}}(h^{(2)}) \otimes 
\cdots \otimes  V_{\lambda ^{(M)}}(h^{(M)})
\end{equation}
where we assume that $h^{(r)}-h^{(s)} \not\in \zint$ for all $r \neq s$.

Let us  set $\rho _{0}(u)=1$ and for $m=1 ,\dots ,n$ let us  define
\begin{equation}
\rho _{m}(u) = \prod _{s=1}^{M} \prod _{i=1}^{m} (u-i+1+h^{(s)}),
\end{equation}
and 
\begin{eqnarray}
& a_m(u)=\rho _m(u)A_m(u) \; & m=0, \cdots ,n \; ,\\
& b_m(u)=\rho _m(u)B_m(u) \; & m=1, \cdots ,n-1 \; ,  \\
& c_m(u)=\rho _m(u)C_m(u) \; & m=1, \cdots ,n-1 \; .
\end{eqnarray}
Then $a_m(u)$, $b_m(u)$, and $c_m(u)$
are polynomials in $u$, 

Let us fix a set of Gelfand-Zetlin schemes 
\begin{equation}
\Lambda ^{(s)} = ( \lambda _{m,i} ^{(s)} | 1 \leq i \leq m \leq n)
 \in {\cal T}_{\l ^{(s)}} \; \; \; (s=1, \dots ,M),
\end{equation}
and define the following polynomials for $m=0, \cdots ,n$. 
\begin{equation}
\varpi _{m, \Lambda ^{(1)}, \dots ,\Lambda ^{(M)}}(u)=
\prod _{s=1}^{M} \prod _{i=1}^{m}(u+\l ^{(s)}_{m,i}-i+1+h^{(s)}).
\end{equation}
Note that all the zeroes of
the $m$--th polynomial
\begin{equation}
\nu_{m,i}^{(s)}=i-\l _{m,i}^{(s)}-1-h^{(s)},
\end{equation}
are pairwise distinct due to our assumption on the parameters
$h^{(1)}, \dots , h^{(M)}$.

For the pairs $(m,m') \; (  1 \leq m' \leq m \leq n)$, we introduce the
ordering,
\begin{equation}
(m,m') \prec (l,l') \; \; \Leftrightarrow \; \; 
m'<l' \mbox{ or } ( \; m'=l' \mbox{ and } m>l ).
\end{equation}
Let $v_{h.w.v} \in W$ be the vector, which is the tensor product of
the highest weight vectors $v^{(s)}_{h.w.v}$ of the Lie algebra
 $\gln $ $(s=1, \cdots ,M)$.
Then consider the following vector in $W$
\begin{equation}
v_{ \Lambda ^{(1)}, \dots ,\Lambda ^{(M)}}  =
\prod_{(m,m')}^\rightarrow\
\left( \prod_{(s,t) \atop{1 \leq t \leq \l ^{(s)}_{n,m'}- \l ^{(s)}_{m,m'}}}
b_m(\nu_{m,m'}^{(s)}-t)
\right) \cdot v_{h.w.v},
\end{equation}
Remark that for each fixed $m$ the elements
$b_m(\nu_{m,m'}^{(s)}-t) \in End(W)$ mutually commute.

Then the following propositions are satisfied. (See \cite{nt2,TU}) 
\begin{prop} {\em \cite{nt2}} 
For every $m=1, \cdots , n$ we have the equality
\begin{equation}
a_m(u) \cdot v_{ \Lambda ^{(1)}, \dots ,\Lambda ^{(M)}}
 =\varpi_{m,\Lambda ^{(1)} , \dots ,\Lambda^{(M)}}(u)
 \cdot v_{ \Lambda ^{(1)}, \dots ,\Lambda ^{(M)}}.
\label{gza}
\end{equation}
\end{prop}
\begin{prop}{\em \cite{nt2}}  
$Y(\gln )$-module $W$ is irreducible
if $h^{(r)}-h^{(s)} \notin \zint $ for all $r \neq s$.
\end{prop}
\begin{prop} \label{bgz}
$v_{ \Lambda ^{(1)}, \dots ,\Lambda ^{(M)}} \; \; 
(\Lambda ^{(r)} \in {\cal S}_{\l^{(r)}}$ for every $r\in\{1, \dots , M\}$) 
form a  base of $W$.
\end{prop}

\section{Yangian in the Spin Calogero Model.}
In this section we recall the definition of the  Yangian action in the spin Calogero model and establish some properties of this action -- in particular the self-adjointness of the operators giving the action of the Gelfand-Zetlin algebra. 

for $\kappa =\pm $ define the Monodromy operator $\hat{T}^
{(\kappa)}_0(u)$ $\in End(\cplxn)\otimes End(\h)[[u^{-1}]]$ by  
\begin{equation}
\hat{T}^{(\kappa)}_0(u) = \sum_{a,b =1}^n E_{a,b}\otimes \hat{T}^{(\kappa)}_{a,b}(u) 
\label{Cmono}
\end{equation}
\vspace{-.15in}
\begin{equation}
:= \left(1 +\frac{P_{0,1}}{u -\kappa \tilde{\Delta }_1}\right)\left(1 + \frac{P_{0,2}}{u-\kappa \tilde{\Delta }_2}\right)\ldots\left(1 + \frac{P_{0,N}}{u -\kappa \tilde{\Delta }_N}\right).
\nonumber
\end{equation}
the $P_{0,i}$ in this definition is the permutation operator of the 0-th and $i$-th tensor factors $\cplxn$ in the tensor product 
\begin{equation}
\underset{0}{\cplxn}\otimes \cz \otimes \underset{1}{\cplxn}\otimes \underset{2}{\cplxn}\otimes \cdots \otimes \underset{N}{\cplxn}  = \underset{0}{\cplxn}\otimes \h.
\end{equation}
The $E_{a,b} \in End(\cplxn)$ is the standard matrix unit in the basis $\{ v_{\ep}\}$.
The operators $ \hat{T}_{a,b}^{(\kappa),(s)} \in End(\h)$ obtained by expanding the Monodromy matrix  $\hat{T}^{(\kappa)}_{a,b}(u)$:    
\begin{equation}
\hat{T}^{(\kappa)}_{a,b}(u) = \delta_{a,b}1  +  \sum_{s\geq 1} u^{-s}  \hat{T}_{a,b}^{(\kappa),(s)}
\end{equation}
satisfy the defining relations (\ref{Ydr}) of the algebra $Y(\gln)$. 
By using 
the relations of the degenerate affine Hecke algebra (\ref{CdH1},\ref{CdH2},\ref{CdH3}) we can show
\begin{equation}
(K_{i,i+1} - \kappa P_{i,i+1}) \hat{T}^{(\kappa)}(u) =
\hat{T}^{(\kappa)}(u) | _{\tilde{\Delta }_{i} 
\leftrightarrow  \tilde{\Delta }_{i+1}}  (K_{i,i+1} - \kappa P_{i,i+1}).
\end{equation}
Then the operators $\hat{T}_{a,b}^{(\kappa),(s)}$ leave the subspace $\hs{\kappa}$ invariant. We will set  
\begin{equation}
T^{(\kappa)}_{a,b}(u) := \left.\hat{T}^{(\kappa)}_{a,b}(u)\right|_{\hs{\kappa}}
\quad \in End(\hs{\kappa})[[u^{-1}]] \qquad (a,b =1,2,\dots,n).  
\label{rest}
\end{equation}

Denote the generating series  which give the action of the Gelfand-Zetlin algebra in the Yangian representation defined by the Monodromy matrix (\ref{Cmono}) by $ A_1^{(\kappa)}(u), A_2^{(\kappa)}(u),\dots , A_n^{(\kappa)}(u) $.
The $A^{(\kappa)}_n(u)$ is just the quantum determinant of the $T^{(\kappa)}_{a,b}(u)$. Hence
\begin{equation}
[A^{(\kappa)}_n(u), T^{(\kappa)}_{a,b}(v)] = 0 \qquad (a,b =1,2,\dots,n).
\end{equation}
The explicit expression for the quantum determinant  \cite{BGHP}:
\begin{equation}
A^{(\kappa)}_n(u) = \prod_{i=1}^N \frac{u + 1 -\kappa \tilde{\Delta }_i}{u -\kappa \tilde{\Delta }_i}.
\end{equation} 
shows that the spin Calogero Hamiltonian (\ref{CH}) is an element in the center of the Yangian action and hence is an element in the Gelfand-Zetlin algebra. 

By using the self-adjointness $\tilde{\Delta }_{i}^{\dagger } =\tilde{\Delta }_{i}$,
we can show the following propositions (cf. \cite{TU} for the proofs).
\begin{prop}
\begin{equation}
{T_{a,b}^{(\kappa)}(u)}^{\dagger} = T_{b,a}^{(\kappa)}(u)  \qquad (\kappa = -,+). 
\end{equation} 
\end{prop}
\begin{prop} \label{sa}
\begin{equation}
{A_{m}^{(\kappa)}(u)}^{\dagger }= A_m^{(\kappa)}(u), \; \; \; {B_{m}^{(\kappa)}(u)}^{\dagger }= C_m^{(\kappa)} (u)
  , \; \; \; {C_m^{(\kappa)}(u)}^{\dagger }= B_m^{(\kappa)} (u) \qquad (\kappa =-,+).
\end{equation} 
\end{prop}

\section{Decomposition of the space of states into irreducible Yangian submodules.} \label{dec}
In this section we construct the decomposition of the space of states of the Spin Calogero model into irreducible submodules of the Yangian action.
The contents of this section is almost the same as \cite{TU} section 5.

 \subsection{Irreducible decomposition of the space of states with respect to the Yangian action. Fermionic case.}
In this subsection we describe the decomposition of the space of states in the fermionic spin Calogero model: $\hs{-}$ into irreducible subrepresentations with respect to the $Y(\gln)$-action (\ref{Cmono}) $(\kappa = -)$. 
Let  $ E^{\mm} := \oplus_{\s\in S^{\mm}}\cplx\Phi_{\s}^{\mm}(A_{1}^{\dagger }, \dots A_{N}^{\dagger }) e^{ -\frac{\omega }{2h} \sum_{j=1}^N x_j ^2 }$ $(\mm \in \MC_N)$. And let  
\begin{equation}
F^{\mm} := ( E^{\mm}\otimes (\Ncplxn))\cap \hs{-}.  
\end{equation}
The (\ref{rest}) implies that the space $ F^{\mm}$ is invariant with respect to the Yangian action defined by (\ref{Cmono}) with $\kappa = -$. And since $\Phi_{\s}^{\mm}(A_{1}^{\dagger }, \dots A_{N}^{\dagger })e^{ -\frac{\omega }{2h} \sum_{j=1}^N x_j ^2} $ $(\mm \in \MC_N, \; \s \in S^{\mm})$ form a base in $\cx$ we have 
\begin{equation}
\hs{-} = \bigoplus_{\mm \in \MC_N} F^{\mm}.
\end{equation}
To describe each of the components $ F^{\mm}$ explicitly we need to make several definitions.

Let $W^{\mm}_{(-)} \subset \Ncplxn $  $(\mm \in \MC_N)$ be defined by 
\begin{equation}
W^{\mm}_{(-)} := \bigcap_{1\leq i \leq N \; \text{s.t.} \; \mlett_i = \mlett_{i+1}} Ker( P_{i,i+1}+ 1). 
\label{wminus}
\end{equation}
Note that $\dim W^{\mm}_{(-)}=0$ unless $\mm \in \MC_N^{(n)}$ where 
\begin{equation}
\MC_N^{(n)}:= \{\; \mm \in \MC_N \;|\; \#\{\;\mlett_k\; |\;\mlett_k = i\;\} \leq n\quad (i\in \zint)\}. 
\end{equation}

For $p \in \{1,2,\dots,n\}$ let $\la$ be the highest weight of the fundamental $\gln$-module:
\begin{equation}
\la = (\underbrace{1,1,\dots,1}_{p},\underbrace{0,0,\dots,0}_{n-p})\qquad ( 1 \leq p \leq n ). 
\end{equation}
For a highest weight of this  form and $h \in \cplx$ denote the corresponding $Y(\gln)$-module $V_{\la}(h)$ by $V_p(h)$. As a linear space the  $V_p(h)$ is realized as the totally antisymmetrized tensor product of $\cplxn$:
\begin{equation}
V_p(h) =  \cap_{i=1}^{p-1} Ker(P_{i,i+1}+1) \quad \subset \quad \otimes^p\cplxn \qquad  (1\leq p \leq n)
\end{equation}
with normalization chosen so that the $\gln$ highest weight vector in  $V_p(h)$
is 
\begin{equation}
{\omega}_p := \sum_{\s \in \Sgroup{p}} (-1)^{l(\s)} v_{\s(1)}\otimes v_{\s(2)} \otimes \cdots \otimes v_{\s(p)} . 
\end{equation}

For an $\mm \in \MC_N^{(n)}$ let $M$ be the number of distinct elements in the sequence 
\begin{equation}
\mm = (\mlett_1,\mlett_2,\dots,\mlett_N). \nonumber
\end{equation}
And let $ p_s $ $( 1\leq p_s \leq n, \quad s=1,2,\dots,M)$ be the multiplicities of the elements in the $\mm$:   
\begin{multline}
\mlett_1=\mlett_2=\cdots =\mlett_{p_1} > \mlett_{1+p_1}= \mlett_{2+p_1}=\cdots =\mlett_{p_2+p_1}> \quad \cdots \\ \cdots > \mlett_{1+ p_{M-1}+\cdots +p_2+p_1}=  \mlett_{2+ p_{M-1}+\cdots +p_2+p_1}=\cdots =\mlett_{p_{M}+\cdots +p_2+p_1 \equiv N}. 
\label{mpf}
\end{multline}
With $\xi^{\mm}_i := \xi^{\mm}_i({\mbox{id}})$ (\ref{xi}) set 
\begin{equation}
h_{\mm}^{(s)} := \xi^{\mm}_{1+p_1+p_2+\cdots+p_{s-1}} \qquad (p_0 :=0,\quad s=1,2,\dots,M ). 
\label{hf}
\end{equation}
Then for the linear space  $W^{\mm}_{(-)}$ (\ref{wminus}) we have 
\begin{equation}
W^{\mm}_{(-)} =  
\begin{cases} {V}_{p_1}(h_{\mm}^{(1)}) \otimes \cdots \otimes {V}_{p_M}(h_{\mm}^{(M)}) \; \subset \; \otimes^N\cplxn  &  \text{ when $ \mm \in \MC_N^{(n)} $,} \\  
                     \qquad \emptyset \qquad  &  \text{ when $ \mm \not\in \MC_N^{(n)} $.}  
\end{cases}
\label{wmtensor}
\end{equation}
 When $ \mm \in \MC_N^{(n)} $ the $W^{\mm}_{(-)}$  is the Yangian module with the Yangian action defined by the coproduct (\ref{copro}).

For any $\s \in S^{\mm}$ (\ref{sm}) define $\check{\real}^{(-)}(\s)$ $\in$ $End(\Ncplxn)$ by the following recursion relation:
\begin{gather}
\check{\real}^{(-)}({\mbox{id}}) := 1, \\
\check{\real}^{(-)}(\s(i,i+1)) := -\check{R}_{i,i+1}\left( \xi_i^{\mm}(\s) - \xi^{\mm}_{i+1}(\s) \right)\check{\real}^{(-)}(\s) \qquad ( \;\mlett_{\s(i)} >\mlett_{\s(i+1)}\;)  
\end{gather}
where the $R$-matrix is given by 
\begin{equation}
\check{R}_{i,i+1}(u) := u^{-1} + P_{i,i+1} .
\label{Rmat}
\end{equation}
Due to the property (\ref{sm}) of the set $ S^{\mm}$ this recursion relation 
is sufficient to define $\check{\real}^{(-)}(\s)$  for all  $\s \in S^{\mm}$. The definition of the $\check{\real}^{(-)}(\s)$ is unambiguous by virtue of the Yang-Baxter equation satisfied by the  $R$-matrix (\ref{Rmat}).

For $\mm \in \MC_N$ define the map $U^{\mm}_{(-)}:$ $ \Ncplxn \rightarrow \h$ by setting for $v \in \Ncplxn$
\begin{equation}
U^{\mm}_{(-)}v := \sum_{\s \in S^{\mm}}\Phi_{\s}^{\mm}(A_{1}^{\dagger}, \dots ,A_{N}^{\dagger })e^{ -\frac{\omega }{2h} \sum_{j=1}^N x_j ^2 } 
\otimes \check{\real}^{(-)}(\s)v. 
\end{equation}

\begin{thm} \label{fisom}
For any $\mm \in \MC_N$ we have
\begin{equation}
U^{\mm}_{(-)}: \: W^{\mm}_{(-)} \rightarrow F^{\mm}. 
\end{equation}
 And the $U^{\mm}_{(-)}$ is an isomorphism of the $Y(\gln)$-modules $W^{\mm}_{(-)}$ and $F^{\mm}$. 
\end{thm}
The proof of this theorem is almost the same as the one given in \cite{TU}
 the Appendix A. (Exchange 
$\Phi_{\s}^{\mm}(z) \leftrightarrow \Phi_{\s}^{\mm}(A_{1}^{\dagger}, \dots ,A_{N}^{\dagger })e^{ -\frac{\omega }{2h} \sum_{j=1}^N x_j ^2 } $)

For now let us notice that from this theorem it follows that the Yangian highest weight vector $\Omega^{(-)}_{\mm}$ in $F^{\mm}$ is given by 
\begin{equation}
\Omega^{(-)}_{\mm} = U^{\mm}_{(-)}\omega_{\mm} = \sum_{\s \in S^{\mm}}
\Phi_{\s}^{\mm }(A_{1}^{\dagger}, \dots ,A_{N}^{\dagger})e^{ -\frac{\omega }{2h} \sum_{j=1}^N x_j ^2 } \otimes \check{\real}^{(-)}(\s)\omega_{\mm} 
\end{equation}
where the $\omega_{\mm}$ is the highest weight vector in $W^{\mm}_{(-)}$:
\begin{equation}
\omega_{\mm} := {\o}_{p_1}\otimes{\o}_{p_2}\otimes \cdots \otimes{\o}_{p_M}. 
\end{equation}

\mbox{} From the corollary 3.9 in \cite{nt2} it follows that the modules $F^{\mm}$ are irreducible if $\alpha \not\in {\Bbb Q}$ since in this case in (\ref{wmtensor}) we have $h^{(s)}_{\mm} -  h^{(r)}_{\mm} \not\in {\Bbb Z}$ when $s \neq r$. 
Using results of \cite{AK} for the Yangian version, we can verify, that the $F^{\mm}$ are irreducible under the weaker condition: $\alpha \in \real\setminus {\Bbb Q}_{\leq 0} $.

\subsection{Irreducible decomposition of the space of states with respect to the Yangian action. Bosonic case.} 
The decomposition of the space of states of the bosonic spin Calogero model: $\hs{+}$ into irreducible sub-representations with respect to the $Y(\gln)$-action (\ref{Cmono}) $(\kappa = +)$ is carried out along the same lines as the one for the fermionic case.  

Let for $\mm \in \MC_N$ the $ E^{\mm} $ be defined as in the previous subsection. And let 
\begin{equation}
B^{\mm} := ( E^{\mm}\otimes (\Ncplxn))\cap \hs{+}.  
\end{equation}
The (\ref{rest}) implies that the space $ B^{\mm}$ is invariant with respect to the Yangian action defined by (\ref{Cmono}) with $\kappa = +$. And since $\Phi_{\s}^{\mm}(A_{1}^{\dagger}, \dots ,A_{N}^{\dagger})e^{ -\frac{\omega }{2h} \sum_{j=1}^N x_j ^2 } $ $(\mm \in \MC_N, \; \s \in S^{\mm})$
  form a base in $\cx$ we have 
\begin{equation}
\hs{+} = \bigoplus_{\mm \in \MC_N} B^{\mm}.
\end{equation}
To describe each of the components $ B^{\mm}$ explicitly we make several definitions analogous to those made in the previous subsection.

Let $W^{\mm}_{(+)} \subset \Ncplxn $  $(\mm \in \MC_N)$ be defined by 
\begin{equation}
W^{\mm}_{(+)} := \bigcap_{1\leq i \leq N \; \text{s.t.} \; \mlett_i = \mlett_{i+1}} Ker( P_{i,i+1} - 1). 
\label{wplus}
\end{equation}

For $p =1,2,\dots $ let $\la$ be the following  $\gln$ highest weight:
\begin{equation}
\la = (p,\underbrace{0,0,\dots,0}_{n-1}). 
\end{equation}
For a highest weight of this  form and $h \in \cplx$ denote the corresponding $Y(\gln)$-module $V_{\la}(h)$ by $V^p(h)$. As a linear space the  $V^p(h)$ is realized as the totally symmetrized tensor product of $
\cplxn$:
\begin{equation}
V^p(h) =  \cap_{i=1}^{p-1} Ker(P_{i,i+1}-1) \quad \subset \quad \otimes^p\cplxn \qquad  (p=1,2,\dots).
\end{equation}
We choose normalization so that the highest weight vector in  $V_p(h)$ is equal to $ v_1^{\otimes p} $ 
As in the fermionic case,  for an $\mm \in \MC_N$ let $M$ be the number of distinct elements in the sequence $\mm = (\mlett_1,\mlett_2,\dots,\mlett_N)$. And let $ p_s $ $(   s=1,2,\dots,M)$ be the multiplicities of the elements in the $\mm$:   
\begin{multline}
\mlett_1=\mlett_2=\cdots =\mlett_{p_1} > \mlett_{1+p_1}= \mlett_{2+p_1}=\cdots =\mlett_{p_2+p_1}> \quad \cdots \\ \cdots > \mlett_{1+ p_{M-1}+\cdots +p_2+p_1}= \mlett_{2+ p_{M-1}+\cdots +p_2+p_1}=\cdots =\mlett_{p_{M}+\cdots +p_2+p_1 \equiv N} . 
\end{multline}
With $\xi^{\mm}_i := \xi^{\mm}_i({\mbox{id}})$ (\ref{xi}) set 
\begin{equation}
h_{\mm}^{(s)} := -\xi^{\mm}_{1+p_1+p_2+\cdots+p_{s-1}} \qquad (p_0 :=0,\quad s=1,2,\dots,M ). 
\label{hb}
\end{equation}
Then for the linear space  $W^{\mm}_{(+)}$ (\ref{wplus}) we have 
\begin{equation}
W^{\mm}_{(+)} =   {V}^{p_1}(h_{\mm}^{(1)})\otimes{V}^{p_2}(h_{\mm}^{(2)})\otimes\ \cdots \otimes 
{V}^{p_M}(h_{\mm}^{(M)}) \; \subset \; \otimes^N\cplxn .
\end{equation}
 The $W^{\mm}_{(+)}$  is the Yangian module with the Yangian action defined by the coproduct (\ref{copro}).

For any $\s \in S^{\mm}$ (\ref{sm}) define $\check{\real}^{(+)}(\s)$ $\in$ $End(\Ncplxn)$ by the following recursion relation:
\begin{gather}
\check{\real}^{(+)}({\mbox{id}}) := 1, \\
\check{\real}^{(+)}(\s(i,i+1)) := \check{R}_{i,i+1}\left( -\xi_i^{\mm}(\s) + \xi^{\mm}_{i+1}(\s) \right)\check{\real}^{(+)}(\s) \qquad ( \;\mlett_{\s(i)} >\mlett_{\s(i+1)}\;)  
\end{gather}
where the $R$-matrix $\check{R}_{i,i+1}(u)$ is given by (\ref{Rmat}).

As in the fermionic case, due to the property (\ref{sm}) of the set $ S^{\mm}$ this recursion relation is sufficient to define $\check{\real}^{(+)}(\s)$  for
 all  $\s \in S^{\mm}$. The definition of the $\check{\real}^{(+)}(\s)$ is unambiguous by virtue of the Yang-Baxter equation satisfied by the  $R$-matrix (\ref{Rmat}).

For $\mm \in \MC_N$ define the map $U^{\mm}_{(+)}:$ $ \Ncplxn \rightarrow \h$ by setting for $v \in \Ncplxn$
\begin{equation}
U^{\mm}_{(+)}v := \sum_{\s \in S^{\mm}}\Phi_{\s}^{\mm}(A_{1}^{\dagger}, \dots ,A_{N}^{\dagger})e^{ -\frac{\omega }{2h} \sum_{j=1}^N x_j ^2 } 
\otimes \check{\real}^{(+)}(\s)v. 
\end{equation}

\begin{thm} \label{bisom}
For any $\mm \in \MC_N$ we have
\begin{equation}U^{\mm}_{(+)}: \: W^{\mm}_{(+)} \rightarrow B^{\mm}. \end{equation}
 And the $U^{\mm}_{(+)}$ is an isomorphism of the $Y(\gln)$-modules $W^{\mm}_{(+)}$ and $B^{\mm}$. 
\end{thm}
We omit the proof of this theorem since it  is a straightforward modification of the proof of the theorem 1 given in \cite{TU}.
\mbox{}From this theorem it follows that the Yangian highest weight vector $\Omega^{(+)}_{\mm}$ in $B^{\mm}$ is given by 
\begin{equation}
\Omega^{(+)}_{\mm} = U^{\mm}_{(+)}v_1^{\otimes N} = \sum_{\s \in S^{\mm}}\Phi_{\s}^{\mm}(A_{1}^{\dagger}, \dots ,A_{N}^{\dagger})
e^{ -\frac{\omega }{2h} \sum_{j=1}^N x_j ^2 } \otimes \check{\real}^{(+)}(\s)  v_1^{\otimes N}.
\end{equation}

\section{Norms of the highest weight vectors in the irreducible Yangian submodules.} 

In this section we will show the concrete expression of the norms ${\sprod{\;\Omega^{(+)}_{\mm}\;}{\;\Omega^{(+)}_{\mm}\;}}_{(+)}$ and ${\sprod{\;\Omega^{(-)}_{\mm}\;}{\;\Omega^{(-)}_{\mm}\;}}_{(-)}$.
The method how to calculate is the same as \cite{TU} section 6. \\
(In the calculation of the norms of the highest weight vectors, we will use the norms of the symmetric generalized Hermite polynomials ${\sprod{\; H_{\mm}(x )e^{ -\frac{\omega }{2h} \sum_{j=1}^N x_j ^2 }\;}{\; H_{\mm}(x)\: e^{ -\frac{\omega }{2h} \sum_{j=1}^N x_j ^2 }\;}}_{c}$
instead of ${\sprod{\;J^{\mm}(z)\;}{\; J^{\mm}(z)\;}}_{J}$.)

\begin{prop} \label{nhwv}
(Bosonic cases)

For $\mm \in \MC_N$ we have
\begin{equation} 
{\sprod{\:\Omega^{(+)}_{\mm}\:}{\:\Omega^{(+)}_{\mm}\:}}_{(+)} = 
(2 \omega )^{|\mm |} {\sprod{\:H^{\mm}(x)\: e^{ -\frac{\omega }{2h} \sum_{j=1}^N x_j ^2 }\:}{\:H^{\mm}(x)\: e^{ -\frac{\omega }{2h} \sum_{j=1}^N x_j ^2 }\:}}_c
\end{equation}
where the norm ${\sprod{\:H^{\mm}(x)\: e^{ -\frac{\omega }{2h} \sum_{j=1}^N x_j ^2 }\:}{\:H^{\mm}(x)\: e^{ -\frac{\omega }{2h} \sum_{j=1}^N x_j ^2 }\:}}_c $ of the symmetric generalized Hermite polynomial is given by the formula (\ref{normsh}).
\\ \mbox{} \\ 
(Fermionic cases)

For $\mm \in \MC^{(n)}_N$ we have 
\begin{multline}
{\sprod{\:\Omega^{(-)}_{\mm}\:}{\:\Omega^{(-)}_{\mm}\:} }_{(-)} =  
\prod_{1\leq s < t \leq M}\frac{\Gamma\left(\:\frac{h_{\mm}^{(s)}-h_{\mm}^{(t)}}{\alpha} + \frac{p_t}{\alpha} + 1\:\right) \:\Gamma\left(\:\frac{h_{\mm}^{(s)}-h_{\mm}^{(t)}}{\alpha} - \frac{p_s}{\alpha} \:\right)}{\Gamma\left(\:\frac{h_{\mm}^{(s)}-h_{\mm}^{(t)}}{\alpha} + \frac{p_t-p_s}{\alpha} + \theta(p_s\leq p_t)\:\right)\:\Gamma\left(\:\frac{h_{\mm}^{(s)}-h_{\mm}^{(t)}}{\alpha} +  \theta(p_s > p_t)\:\right)}  \\
\prod_{i=1}^N \Gamma (m_i +1+\frac{N-i}{\alpha }) \: \cdot \:
\prod_{s=1}^M \frac{\Gamma\left(\:\frac{p_s}{\alpha}+1\:\right)}{\left\{\Gamma\left(\:\frac{1}{\alpha}+1\:\right)\right\}^{p_s}}\: \cdot \: 
 h  ^{-|\mm |}
\left( \frac{\omega }{2h} \right) ^{ \frac{N(N-1)}{2\alpha }}
 \left( \frac{\omega \pi}{h } \right) ^{\frac{N}{2}} 
N ! ,
\end{multline} 
where 
\begin{equation}
\theta(x):= \begin{cases} \; 1 \; &  \text{{\em when $x$ is true,} } \\ \; 0 \; &  \text{{\em when $x$ is false.} } \end{cases}
\end{equation}
\end{prop}

\section{Eigenbases of the Gelfand-Zetlin algebra in the irreducible Yangian submodules and norms of the eigenvectors.} 
In this section we construct eigenbases of the operator-valued series $A_1^{(\kappa)}(u), A_2^{(\kappa)}(u), \dots A_n^{(\kappa)}(u)$ within each of the irreducible $Y(\gln)$-submodules $F^{\mm}$ $(\mm \in \MC^{(n)}_N)$ ( $\kappa = -1$ --fermionic case ) and $B^{\mm}$ $(\mm \in \MC_N)$ ( $\kappa = 1$ -- bosonic case ), and compute norms of the eigenvectors that form these eigenbases.
 
Due to the isomorphisms given by the theorems \ref{fisom} and \ref{bisom} the 
construction of the eigenbases is carried out by a straightforward application of the results of Nazarov and Tarasov \cite{nt2}.

Let us fix a partition $\mm = (\mlett_1,\mlett_2,\dots,\mlett_N)  \in \MC_N$ and let for $\kappa=-1$ $\mm \in \MC^{(n)}_N \subset \MC_N$.
Define the number $p_{s}$ to be the multiplicities of the elements in the $\mm $ (see (\ref{mpf})).
 In the fermionic case $\mm \in \MC^{(n)}_N$ and $  p_s \in \{1,2,\dots,n\} $ $( s=1,2,\dots,M)$.

For $ p\in \{1,2,\dots,n\} $ let $ {\cal S}^{(-)}_p $ denote the set of all Gelfand-Zetlin schemes $ \Lambda $ that are associated with the irreducible $\gln$-module with  the highest weight (cf. sec. \ref{dec})
\begin{equation}
 (\underbrace{1,1,\dots,1}_{p},\underbrace{0,0,\dots,0}_{n-p}).
\end{equation}
An element of $ {\cal S}^{(-)}_p $ is an array of the form
\begin{eqnarray}
& \l _{n,1} \; \; \l _{n,2} \; \cdots \cdots \cdots \cdots \cdots \; \; 
 \l _{n,n} &  \label{fgzs} \\
& \l _{n-1,1} \; \;   \cdots \cdots \; \; \l _{n-1,n-1} &
\nonumber \\
& \ddots \; \; \cdots \cdots \; \; \; \; \; \; \;  & \nonumber \\
& \l _{2,1} \; \; \l _{2,2} & \nonumber \\
& \l _{1,1} & \nonumber    
\end{eqnarray}
where 
\begin{gather}
(\l_{m,1},\l_{m,2}, \dots ,\l_{m,m}) = (\underbrace{1,1,\dots,1}_{l_{m}},\underbrace{0,0,\dots,0}_{m-l_{m}}) \qquad ( m=1,2,\dots,n),  \\ 
l_n = p \nonumber  
\end{gather}
and 
\begin{equation}
\text{either $ \quad l_{m} = l_{m+1} \quad $  or  $  \quad l_{m} = l_{m+1}-1 \quad $ } \qquad ( m=1,2,\dots,n-1).
\end{equation}

For $ p\in {\Bbb N} $ let $ {\cal S}^{(+)}_p $ denote the set of all Gelfand-Zetlin schemes $ \Lambda $ that are associated with the irreducible $\gln$-module with the highest weight (cf. sec. \ref{dec})
\begin{equation}
 ({p},\underbrace{0,0,\dots,0}_{n-1}).
\end{equation}
An element of  $ {\cal S}^{(+)}_p $ is a Gelfand-Zetlin scheme of the form 
\begin{eqnarray}
& \alpha _{n} \; \; 0  \; \cdots \cdots \cdots \cdots \cdots \; \; 
 0 &  \label{bgzs} \\
& \alpha _{n-1} \; \;  0 \; \;  \cdots \cdots \; \; 0 &
\nonumber \\
& \ddots \; \; \cdots \cdots \; \; \; \; \; \; \;  & \nonumber \\
& \alpha _{2} \; \; 0 & \nonumber \\
& \; \; \alpha _{1} & \nonumber 
\end{eqnarray}
where 
\begin{gather}
\alpha_{m} \leq \alpha_{m+1} \qquad ( m=1,2,\dots,n-1), \\
\alpha_n = p. \nonumber  
\end{gather}

Now let us define the following operator-valued series:\\
For the bosonic case  set
\begin{equation}
a_{m}^{(+)}(u)= A^{(+)}_m(u), \; \;  b_{m}^{(+)}(u)= B^{(+)}_m(u), \; \; 
c_{m}^{(+)}(u)=  C^{(+)}_m(u). 
\end{equation}
And for the fermionic case   set
\begin{gather}
a_{m}^{(-)}(u)= \Delta (u) A^{(-)}_m(u), \; \;  
b_{m}^{(-)}(u)= \Delta (u) B^{(-)}_m(u), \; \;  
c_{m}^{(-)}(u)= \Delta (u) C^{(-)}_m(u),  \nonumber 
\end{gather}
where $\Delta (u) = \prod _{i=1}^{N} (u+\hat{\Delta }_i)$.
Then from the proposition \ref{sa} it follows that 
\begin{equation} 
a_{m}^{(\kappa)}(u) ^{\dagger }= a_m^{(\kappa)} (u), \; \; \; 
b_{m}^{(\kappa)}(u) ^{\dagger }= c_m ^{(\kappa)}(u)  , \; \; \;  
c_{m}^{(\kappa)}(u) ^{\dagger }= b_m ^{(\kappa)}(u),  \; \; \; \kappa=-,+ .
\end{equation} 

For a collection of Gelfand-Zetlin schemes $\Lambda ^{(1)}, \dots ,\Lambda ^{(M)}$ such that $ \Lambda ^{(s)} \in {\cal S}^{(\kappa)}_{p_s} $ $ (s=1,2,\dots,M)$ define the following vector (cf. sec. \ref{ygh}):
\begin{align}
& v^{(\kappa)}_{ \Lambda ^{(1)}, \dots ,\Lambda ^{(M)}}  =
\prod_{(m,m')}^\rightarrow\
\left( \prod_{(s,t) \atop{1 \leq t \leq \l ^{(s)}_{n,m'}- \l ^{(s)}_{m,m'}}}
b_m^{(\kappa)}({\nu }_{m,m'}^{(s)}-t)
\right) \cdot \Omega^{(\kappa)}_{\mm}, \\
& v^{(\kappa)}_{ \Lambda ^{(1)}, \dots ,\Lambda ^{(M)}} \in 
\begin{cases} F^{\mm} 
 &  (\kappa = - ), \\ B^{\mm} &  (\kappa = + ). \end{cases} 
\end{align}
Here 
\begin{equation}
{\nu }_{m,m'}^{(s)} = m'-\l _{m,m'}^{(s)}-1 - h^{(s)}_{\mm} 
\end{equation}
and the $h^{(s)}_{\mm}$ are defined by (\ref{hf}) (the fermionic case) and (\ref{hb}) (the bosonic cases).
\mbox{} From  the proposition \ref{bgz} and the theorems \ref{fisom},\ref{bisom} it follows that  the set 
\begin{equation}
  \{ v^{(\kappa)}_{ \Lambda ^{(1)}, \dots ,\Lambda ^{(M)}} \; | \;  
 \; \Lambda ^{(s)}
 \in {\cal S}^{(\kappa)}_{p_s}\quad (s=1,2,\dots,M) \} 
\end{equation}
is a base of  $F^{\mm }$  ( resp. $B^{\mm}$) when $ \kappa = - $ ( resp. $ +$). 
Due to the proposition \ref{gza} this is an eigenbase of the operators generating
 the Gelfand-Zetlin algebra:  
\begin{align}
& A_m^{(\kappa)}(u)\, v^{(\kappa)}_{ \Lambda ^{(1)}, \dots ,\Lambda ^{(M)}} = {\cal A}_m^{(\kappa)}(u;\mm)_{ \Lambda ^{(1)}, \dots ,\Lambda ^{(M)}} v^{(\kappa)}
_{ \Lambda ^{(1)}, \dots ,\Lambda ^{(M)}}, \qquad (m=1,2,\dots,n)\\
\intertext{where the eigenvalues are defined by :}
& {\cal A}_m^{(-)}(u;\mm)_{ \Lambda ^{(1)}, \dots ,\Lambda ^{(M)}} = \prod_{s=1}^M \frac{u + 1 + h^{(s)}_{\mm}} {u + 1 + h^{(s)}_{\mm} - l_m^{(s)}},  \qquad (\Lambda^{(s)} \in {\cal S}^{(-)}_{p_s}); \\ 
&{\cal A}_m^{(+)}(u;\mm)_{ \Lambda ^{(1)}, \dots ,\Lambda ^{(M)}} = \prod_{s=1}^M \frac{u + h^{(s)}_{\mm} + \alpha_m^{(s)}} {u + h^{(s)}_{\mm}},\:\quad\qquad (\Lambda^{(s)} \in {\cal S}^{(+)}_{p_s}).
\end{align}

Since ${\sprod{\; \Phi ^{\mm }_{\s }(A_{1}^{\dagger}, \dots ,A_{N}^{\dagger})e^{ -\frac{\omega }{2h} \sum_{j=1}^N x_j ^2} \;}{\; \Phi ^{\nn}_{\tau }(A_{1}^{\dagger}, \dots ,A_{N}^{\dagger})e^{ -\frac{\omega }{2h} \sum_{j=1}^N x_j ^2} \;}}_c =0 $ when $ \mm \neq \nn $,  the subspaces $F^{\mm }$ (resp. $B^{\mm }$)
 are pairwise orthogonal. 

For $\alpha > 0$ one can verify, that the data $\mm \in \MC_N$, $( \Lambda ^{(1)},\Lambda ^{(2)} \dots ,\Lambda ^{(M)})$ $(\Lambda ^{(s)}  \in  {\cal S}^{(\kappa)}_{p_s})$ are uniquely restored from the collection of rational functions
\begin{equation}
{\cal A}_1^{(\kappa)}(u;\mm)_{ \Lambda ^{(1)}, \dots ,\Lambda ^{(M)}},{\cal A}_2^{(\kappa)}(u;\mm)_{ \Lambda ^{(1)}, \dots ,\Lambda ^{(M)}},\dots,{\cal A}_n^{(\kappa)}(u;\mm)_{ \Lambda ^{(1)}, \dots ,\Lambda ^{(M)}}.
\end{equation}
That is the joint spectrum  of eigenvalues of the Gelfand-Zetlin algebra is simple. Since $A_m ^{(\kappa)}(u)$ are self-adjoint, we obtain 
\begin{prop}
For $\mm \in \MC_N^{(n)} $  ( resp.  $\mm \in \MC_N $ )  the set 
\begin{equation}
  \{ v^{(\kappa)}_{ \Lambda ^{(1)}, \dots ,\Lambda ^{(M)}} \; | \;  
 \; \Lambda ^{(s)}
 \in {\cal S}^{(\kappa)}_{p_s}\quad (s=1,2,\dots,M) \}
\end{equation}
with $\kappa = - $ (resp. $\kappa = + $)  is an  orthogonal base of $F^{\mm }$  (resp. $B^{\mm }$).
\end{prop}

The norms of the eigenvectors  $v^{(\kappa)}_{ \Lambda ^{(1)},\dots ,\Lambda ^{(M)}}$ are as follows,

\begin{prop} 

(Bosonic case) 

Let $\mm \in \MC_N$ and $ \Lambda^{(s)} \in {\cal S}^{(+)}_{p_s} $ $( s=1,2,\dots,M).$ If we write a Gelfand-Zetlin scheme $ \Lambda^{(s)}$ as in (\ref{bgzs}): 
\begin{eqnarray}
& \alpha _{n}^{(s)} \; \; 0  \; \cdots \cdots \cdots \cdots \cdots \; \; 
 0 &  \\
& \alpha _{n-1}^{(s)} \; \;  0 \; \;  \cdots \cdots \; \; 0 &
\nonumber \\
\Lambda^{(s)} = & \ddots \; \; \cdots \cdots \; \; \; \; \; \; \;  & \nonumber \\
& \alpha _{2}^{(s)} \; \; 0 & \nonumber \\
& \; \; \alpha _{1}^{(s)} & \nonumber
\end{eqnarray}
then the norm of the vector $v^{(+)}_{ \Lambda ^{(1)},\dots ,\Lambda ^{(M)}}$ is
 
\begin{multline}
{\langle  v^{(+)}_{ \Lambda ^{(1)}, \dots ,\Lambda ^{(M)}},
v^{(+)}_{ \Lambda ^{(1)}, \dots ,\Lambda ^{(M)}} \rangle}_{(+)}
=  
\qquad  \qquad {\sprod{\:\Omega_{\mm}^{(+)}\:}{\:\Omega_{\mm}^{(+)}\:}}_{(+)}
 \cdot \\
\cdot  \prod _{1 \leq m \leq n}
\left\{
\prod _{1 \leq s \leq M}
\frac{(\alpha _{n}^{(s)} - \alpha _{m}^{(s)})!
(\alpha _{n}^{(s)} - \alpha _{m-1}^{(s)})!
(\alpha _{m}^{(s)}!)^2}{
(\alpha _{m}^{(s)} - \alpha _{m-1}^{(s)})!
(\alpha _{n}^{(s)}!)^2}
\right.
  \\
\left\{ 
\prod _{(s,s') \atop{s \neq s'} } 
\prod _{a=\alpha _{m}^{(s)}}^{\alpha _{n}^{(s)} -1}
\frac{
(-a+\alpha _{n}^{(s')} +h_{\mm}^{(s')}-h_{\mm}^{(s)} )
(-1-a+\alpha _{m-1}^{(s')} +h_{\mm}^{(s')}-h_{\mm}^{(s)} )}{
(-1-a +h_{\mm}^{(s')}-h_{\mm}^{(s)} )^2}
\right\}
 \\
\left. 
\prod _{(s,s') \atop{s < s'} }
\frac{(\alpha _{n}^{(s')}-\alpha _{n}^{(s)} +h_{\mm}^{(s')}-h_{\mm}^{(s)})}{
(\alpha _{m}^{(s')}-\alpha _{m}^{(s)} +h_{\mm}^{(s')}-h_{\mm}^{(s)})}
\right\}
 \\  \mbox{} 
\end{multline}

where the $h_{\mm}^{(s)}$ are defined by (\ref{hb}) with $\kappa = +$.
\\ \mbox{} \\ 
(Fermionic case)

Let $\mm \in \MC_N^{(n)}$ and $ \Lambda^{(s)} \in {\cal S}^{(-)}_{p_s} $ $( s=1,2,\dots,M).$
As in (\ref{fgzs}) define $l_m^{(s)}$ associated with the Gelfand-Zetlin scheme $\Lambda^{(s)}$ by the conditions $\l _{m,l_m^{(s)}}^{(s)} =1$ and
 $\l _{m,l_m^{(s)}+1}^{(s)} =0$. Then the norm of the vector $ v^{(-)}_{ \Lambda ^{(1)}, \dots ,\Lambda ^{(M)}}$ is 
\begin{multline}
{\langle  v^{(-)}_{ \Lambda ^{(1)}, \dots ,\Lambda ^{(M)}},
v^{(-)}_{ \Lambda ^{(1)}, \dots ,\Lambda ^{(M)}} \rangle}_{(-)}
= \qquad \qquad {\sprod{\:\Omega_{\mm}^{(-)}\:}{\:\Omega_{\mm}^{(-)}\:}}_{(-)} \cdot \\
\cdot \left\{
\prod _{1 \leq s \leq M}
\prod_{(m,m')  \atop{\lambda_{m,m'}^{(s)} \neq \lambda _{n,m'}^{(s)}}} 
(m'-1)!^{2}(p_s +1-m')!^{2}
\right\} \\ \mbox{} 
\end{multline}
\begin{equation}
\left\{
\prod _{(s,s') \atop{s < s'} }
\prod_{(m,m')  \atop{\lambda_{m,m'}^{(s)} \neq \l _{n,m'}^{(s)}
\atop{\lambda_{m,m'}^{(s')} \neq \l _{n,m'}^{(s')}}}} 
\frac{\prod_{j=0}^{p_s}(m'-j-1+h_{\mm}^{(s')}-h_{\mm}^{(s)})^{2}
\prod_{j=0}^{p_{s'}}(m'-j-1+h_{\mm}^{(s)}-h_{\mm}^{(s')})^{2}}{
(h_{\mm}^{(s)}-h_{\mm}^{(s')})^{4}}
\right\}
\nonumber
\end{equation}
\begin{equation}
\left\{
\prod _{(s,s') \atop{s \neq s'} }
\prod_{(m,m')  \atop{\lambda_{m,m'}^{(s)} \neq \l _{n,m'}^{(s)}
\atop{\lambda_{m,m'}^{(s')} = \l _{n,m'}^{(s')}}}} 
\frac{(m'-l_{m}^{(s')}+h_{\mm}^{(s')}-h_{\mm}^{(s)})
\prod_{j=0}^{p_{s}}(m'-j-1+h_{\mm}^{(s')}-h_{\mm}^{(s)})^{2}}{
(m'-1-l_{m-1}^{(s')}+h_{\mm}^{(s')}-h_{\mm}^{(s)})
(m'-1-l_{m}^{(s')}+h_{\mm}^{(s')}-h_{\mm}^{(s)})
(m'-l_{m+1}^{(s')}+h_{\mm}^{(s')}-h_{\mm}^{(s)})}
\right\}
\nonumber
\end{equation}
where $h_{\mm}^{(s)}$ are defined by (\ref{hf}) with $\kappa = -$.
In these product formulas the $s$ and $s'$ range from $1$ to $M$ and $ (m,m')$ ( $ n \geq m \geq m' \geq 1 $ ) are coordinates of points in a Gelfand-Zeltin scheme of $\gln$.  
\end{prop}

The proof is the same as the proof of the proposition 14 \cite{TU}.
(Also see appendix B \cite{TU})

Together with the proposition \ref{nhwv}, this proposition gives the norm formulas for the orthogonal eigenbasis of the spin Calogero model.


\begin{thebibliography}{99}
\bibitem{AK} T. Akasaka, and M. Kashiwara,: To be published.
\bibitem{BF1} 
T. H. Baker, and P. J. Forrester,: The Calogero-Sutherland model and generalized classical polynomials, Preprint RIMS-1094 (1996). (solv-int/9608004)
\bibitem{BF2} 
T. H. Baker, and P. J. Forrester,: The Calogero-Sutherland model and polynomials with prescribed symmetry, Preprint (1996). (solv-int/9609010)
\bibitem{BF3} 
T. H. Baker, and P. J. Forrester,: Non-Symmetric Jack Polynomials and Integral Kernels, Preprint (1996). (q-alg/9612003)
\bibitem{BHW}
D. Bernard, K. Hikami, and M. Wadati,: New Developments of Integrable Systems and Long-Ranged Interaction Models, ed. M. L. ge and Y. S. Wu (World Scientific, Singapore) 1 (1995).
\bibitem{BGHP}
D. Bernard, M. Gaudin, F. D. M. Haldane, and V. Pasquier,: J. Phys., {\bf A26} 5219 (1993).
\bibitem{Calogero}
F. Calogero,: J. Math. Phys. {\bf 12} 419 (1971). 
\bibitem{C} I. V. Cherednik,: Duke Math. J. {\bf 54} 563 (1987).
\bibitem{Cherednik2} 
I. V. Cherednik,: Annals Math., {\bf 141} 191 (1995); Non-symmetric Macdonald Polynomials, IMRN {\bf 10} 483  (1995).
\bibitem{Drinfeld}
V. G. Drinfeld,: Sov. Math. Dokl. {\bf 36} 212 (1988); ``Quantum Groups'' in Proceedings of the International Congress of Mathematicians, Amer. Math. Soc., Providence, RI, 798 (1987). 
\bibitem{Dunkl}
C. F. Dunkl,: Trans. Amer. Math. Soc. {\bf 311} 167 (1989).
\bibitem{gz}  I. M. Gelfand, and  M. L. Zetlin,: Dokl. Akad. Nauk SSSR, {\bf 71} 825 (1950).
\bibitem{Kakei}
S. Kakei,: Common Algebraic Structure for the Calogero-Sutherland Models. Preprint (1996). (solv-int/9608009)
\bibitem{Knop}
F. Knop, and S. Sahi,: A recursion and combinatorial formula for Jack polynomials, Preprint (1996). (q-alg/9610016)
\bibitem{Macdonald1} 
I. G. Macdonald,: Affine Hecke algebra and Orthogonal Polynomials, S\'{e}minaire Bourbaki, {\bf 47} No. 797, 1 (1995).
\bibitem{Macdonaldbook} 
I. G. Macdonald,: ``Symmetric functions and Hall polynomials'', 2-nd ed., Clarendon Press (1995).
\bibitem{nt1} 
M. Nazarov, and V. Tarasov,: Publ. Res. Inst. Math. Sci. {\bf 30} 459 (1994).
\bibitem{nt2} 
M. Nazarov, and V. Tarasov,: Representations of Yangians with Gelfand-Zetlin bases, { Preprint} UWS-MRRS-94-148 (1994).
\bibitem{Stanley}
R. P. Stanley,: Adv. Math., {\bf 77} 76 (1989). 
\bibitem{Sutherland}
B. Sutherland,: J. Math. Phys. {\bf 12} 246, 251 (1971); Phys. Rev. {\bf A4} 2019 (1971); ibid. {\bf A5} 1372 (1972).
\bibitem{TU}
K. Takemura,and D. Uglov,: The Orthogonal Eigenbasis and Norms of Eigenvectors in the Spin Calogero-Sutherland Model, { Preprint} RIMS-1114 (1996). (solv-int/9611006)
\bibitem{Ujino}
H. Ujino, and M. Wadati,: J. Phys. Soc. Jpn. {\bf 65} 2423 (1996).



\end{thebibliography}
\end{document}